\documentclass[prd,aps,nofootinbib,notitlepage%,showpacs,preprintnumbers
]{revtex4}
\usepackage{graphicx,epsf,amsmath,amsfonts,amssymb,amsbsy,textcomp}
\usepackage{epsfig}
\newcommand{\vev}[1]{\langle#1\rangle}
\newcommand{\mat}{\left ( \begin{array}}
\newcommand{\emat}{\end{array} \right )}
\newcommand{\vect}{\left ( \begin{array}{c}}
\newcommand{\evect}{\end{array} \right )}

\begin{document}

%\hfill HU-EP-11/11

\title{%1) Duality properties of dense quark matter with chiral and isospin asymmetries: the two-color NJL model consideration\\
%2) Dualities between some condensed matter phenomena considered in two-color NJL model\\
%3) Two-color approach to the duality properties of chiral and isospin asymmetric dense quark matter: the NJL model consideration\\
Influence of chiral asymmetry on phase structure of the two-color quark matter
%The dual properties of chiral and isospin asymmetric dense quark matter formed of two-color quarks

%The dual properties of two-color quark matter
}
\author{T. G. Khunjua $^{1}$, K. G. Klimenko $^{2}$, and R. N. Zhokhov $^{2,3}$ }
%, and V.C. Zhukovsky $^{2)}$}
%\vspace{2cm}

%\affiliation{%$^{1)}$ Institute of Physics,
%Humboldt-University Berlin, 12489 Berlin, Germany}
%\affiliation{$^{1)}$ Faculty of Physics, Moscow State University,
%119991, Moscow, Russia}
\affiliation{$^{1}$ The University of Georgia, GE-0171 Tbilisi, Georgia}
%\affiliation{$^{2)}$ Department of Theoretical Physics, A. Razmadze Mathematical Institute, I. Javakhishvili Tbilisi State University, GE-0177 Tbilisi, Georgia}
%\affiliation{$^{1)}$ A. Razmadze Mathematical Institute, Georgian Academy of Sciences, 380093 Tbilisi, Republic of Georgia}
\affiliation{$^{2}$ State Research Center
of Russian Federation -- Institute for High Energy Physics,
NRC "Kurchatov Institute", 142281 Protvino, Moscow Region, Russia}
%\affiliation{$^{3)}$ University ``Dubna`` (Protvino branch), 142281, Protvino, Moscow Region, Russia}
\affiliation{$^{3}$  Pushkov Institute of Terrestrial Magnetism, Ionosphere and Radiowave Propagation (IZMIRAN),
108840 Troitsk, Moscow, Russia}

\begin{abstract}
In this paper the influence of chiral chemical potential $\mu_5$ on the phenomenon of diquark condensation and phase 
structure of dense quark matter altogether is contemplated in the framework of effective two-color and two-flavor 
Nambu--Jona-Lasinio model. The nonzero values of baryon $\mu_B$, isospin $\mu_I$ and chiral isospin $\mu_{I5}$
chemical potentials are also taken into account. We show that the duality relations between diquark condensation,
charged pion condensation and chiral symmetry breaking phenomena, found in the case of zero $\mu_5$, 
are also valid for any value of $\mu_5\neq0$. Hence the dualities seem to be a real fundamental feature of the 
phase diagram of two color quark matter, and besides they are a very powerful tool of studying the phase structure.
In terms of dualities and the influence on the phase diagram chiral imbalance $\mu_5$ stands alone from other 
chemical potentials, it does not transform under the duality transformations.
Two regimes in the phase diagram could be clearly discerned, the first one, rather small or moderate values of 
chemical potentials $\mu_B$, $\mu_I$ and $\mu_{I5}$, where the picture is rather concise and elegant: each of the above
phenomena is in one-to-one correspondence with some chemical potential ($\mu_{I5}$ induces chiral symmetry breaking, $\mu_I$ propels charged pion condensation
and $\mu_B$ leads to diquark condensation). In this regime chiral chemical potential $\mu_5$ does not break this 
correspondence and does not take part in the determination of the phase,
its role is quite curious, it has a property of being universal catalyzer, i.e. it {\it catalyzes/enhances} all the 
phenomena on equal footing. The feature of chiral imbalance $\mu_5$ to catalyze chiral symmetry breaking phenomenon 
was known before but its catalytic properties appeared to be more universal and it can catalyze also diquark and 
charged pion condensation. These happen due to the fact that phase diagram possesses high symmetry that in turn is 
a consequence of the dual properties. In the second regime, when several other chemical potentials reach rather 
large values, one could observe a rather complicated and rich phase structure, and chiral chemical potential can
be a factor that not so much catalyzes as {\it triggers} %new phases 
%be witnessed to not just catalyze but {\it trigger}
rather peculiar phases. 
For example, it can lead to the 
formation of diquark condensation at zero baryon chemical potential. 
%Furthermore, in the transitory regime, on the borderline between these regimes, especially rich phase structure is observed and a series of first order phase transitions between all the possible phases of the system take place. %Chiral chemical potential exhibits rather peculiar property, it was called chameleon property, depending on conditions it can mimic the influence of any other chemical potential so it was called chameleon property.
Furthermore, chiral chemical potential exhibits rather peculiar property, depending on conditions it can mimic the influence of any other chemical potential so it was called chameleon property.
%The phase structure possesses very high symmetry due to duality properties and it constrains the influence of $\mu_5$ on the phase diagram. The property of being universal catalyzer and the chameleon nature is a consequence of this symmetry.
%Chiral imbalance has been shown to have a property of mimicking any chemical potential and trigger the corresponding symmetry breaking pattern. 
%Furthermore, in the transitory regime, on the borderline between these regimes, especially rich phase structure is observed and a series of first order phase transitions between all the possible phases of the system take place. %Chiral chemical potential exhibits rather peculiar property, it was called chameleon property, depending on conditions it can mimic the influence of any other chemical potential so it was called chameleon property.
\end{abstract}

\maketitle

\section{Introduction}
It is well known that quantum chromodynamics (QCD) is the theory of hot and dense strongly interacting matter.
And the great interest are properties and phase diagram of dense baryon (quark) matter, which may be realized in heavy-ion collision experiments or inside compact stars \cite{fukushima}. However, at the corresponding values of temperature and baryon density, the QCD interaction constant is quite large. Therefore, using the usual perturbation method, it is impossible to obtain an adequate picture of the phenomena of dense matter. In this case different effective low-energy QCD-like models, among which are the well-known Nambu--Jona-Lasinio (NJL) type models \cite{njl,buballa}, etc, can be used to describe the corresponding parts of the QCD phase diagram. Alternatively, one can apply numerical lattice Monte Carlo simulation methods (see, e.g., Ref. \cite{lattice}). But due to a notorious sign problem of the quark determinant, the first-principle lattice approach in the three-color QCD is limited by the systems with zero baryon chemical potential $\mu_B$, i.e. when baryon density of the system is equal to zero. \footnote{Note that throughout the paper we deal only with quarks belonging to the fundamental representation of the color $SU(N_c)$ group. If they are in the adjoint representation of the $SU(N_c=3)$, then the sign problem is absent \cite{son2}.}

In contrast, in the QCD with an even flavor number $N_f$ (below we consider the two-flavor case, i.e. $N_f=2$) of two-colored quarks the
sign problem of the quark determinant is absent at $\mu_B\ne 0$. Moreover, this theory shares such an important aspect of real ($N_c=3$)-color QCD as
spontaneous chiral symmetry breaking at low temperatures. Or the low-energy excitations in both theories consist of color-singlet hadrons,
etc, i.e. the 2-color QCD is a good basis to test various ideas of the 3-color QCD at $\mu_B\ne 0$.  So a lot of investigations of the phase
diagram of two-color QCD have been carried out at $\mu_B\ne 0$ both using the lattice approach and within the framework of some low-energy
effective models \cite{kogut,son2,weise,ramos, andersen3,brauner1,andersen2,imai,adhikari,chao,Bornyakov:2020kyz,khunjua,Furusawa:2020qdz}.
However, for the sake of clarity, it should be noted that there are serious qualitative 
differences between two- and three-color QCD. Namely,
in the first case baryons, including diquarks, are bosons, while at $N_c=3$ all baryons are fermions. In addition, in the massless case the
flavor (chiral) symmetry of the 3-color QCD is $SU(N_f)_L\times SU(N_f)_R\times U(1)$, whereas in the 2-color massless QCD this flavor symmetry is
enhanced to $SU(2N_f)$, and sometimes referred to as the Pauli-Gursey symmetry \cite{kogut}. Moreover, 
as it was recently shown in Ref. \cite{Datta}, the standard model based on two-color QCD 
can have quite different properties than at $N_c=3$, etc.

Strictly speaking, all of the above refers to the description of the properties of hypothetical quark matter, which is characterized by only
one, baryonic, chemical potential $\mu_B$. However, dense baryonic matter, which can exist in neutron stars or be even observed in heavy-ion
collision experiments, is characterized by at least one more property. It is isospin asymmetry. And in this case there appears an additional
isospin chemical potential $\mu_I$ in the system. Moreover, since nuclear matter is usually under the influence of extremely strong external
magnetic fields, chiral asymmetry of the medium can also be observed.
%%%%%%%%%%%
In the most general case this phenomenon is described by two chemical
potentials, chiral $\mu_5$ and chiral isospin $\mu_{I5}$. Thus, the phase structure of real dense quark (baryonic) matter should be described, strictly speaking, in terms of QCD with several chemical potentials. But in this case,  we still have no reliable first-principle computations for finite-density QCD both in the three and two-color quark approaches, despite the fact that we are expecting rich dynamics and its importance inside neutron stars or in heavy-ion experiments. The reason again lies in the sign problem of the quark determinant, which arises in the lattice approach even in the framework of the 2-color QCD when describing dense quark matter with additional chemical potentials.  \footnote{For example, it is shown in Ref. \cite{Furusawa:2020qdz} that the 2-color QCD suffers from the sign problem if both $\mu_B$ and $\mu_I$ are present.} And, therefore, in this case the properties of dense baryonic medium, formed from both 3-color and hypothetical 2-color quarks, are most adequately described only within the framework of low-energy effective models such as NJL models, etc.

Recently, the properties of $N_c=3$ dense quark matter with isospin asymmetry have been studied just in the framework of various NJL-type models
with two quark flavors, where, in particular, it was noted that at $\mu_I >m_\pi$ (here $m_\pi$ is the pion mass) in such a medium a phase
with condensation of charged pions can be observed (for it we use the notation charged PC phase)
\cite{he,abuki,ak,ekkz,Mammarella:2015pxa,Andersen:2018nzq,farias}. Furthermore, if in addition the chiral asymmetry of
quark matter is also taken into account, i.e. $\mu_5\ne 0$ or $\mu_{I5}\ne 0$, then the possibility of appearance of the charged PC
phenomenon in the system is predicted with even greater reliability \cite{zhokhov,kkz2,kkz,kkz1+1}. Notice that in the last papers it was also
shown that at $\mu_I\ne 0$ and $\mu_{I5}\ne 0$, as well as at $\mu_{B}\ne 0$, there is a duality between the phenomena of spontaneous chiral
symmetry breaking (CSB) and condensation of charged pions. It means that on the $(\mu_{B},\mu_{I},\mu_{I5})$-phase diagram of quark matter,
\footnote{Here and below we discuss the properties of dense quark matter formed of $u$ and $d$ quark flavors only.} obtained in the framework
of these simplest $N_c=3$ NJL models, (i) only nontrivial CSB and charged PC phases are present, and (ii) these phases are arranged
symmetrically on the full phase portrait of quark matter (at zero bare quark mass $m_0$ the dual symmetry is exact, but if $m_0\ne 0$ it is an
approximate one \cite{kkz2}). However, since only the simplest, quark-antiquark, of all possible quark interaction channels was taken into
account in the structure of these $N_c=3$ NJL models, the results of Refs. \cite{zhokhov,kkz2,kkz,kkz1+1} can be trusted in the region of
rather low values of baryon densities, i.e. at $\mu_B<1$ GeV, where other phenomena such as color superconductivity, etc are forbidden. And
this is despite the fact that the above mentioned dual symmetry (property) of quark matter at low baryon densities is consistent with calculations, for
example, of the (pseudo)-critical temperature of its crossover transition to the quark gluon plasma phase, made in the lattice approach
(see, e.g., the discussion in Ref. \cite{kkz2}). At higher densities, i.e. at $\mu_B>1$ GeV, in addition to the quark-antiquark, it is necessary to take into account other, for example, diquark, etc, interaction channels, which also participate in the formation of the phase portrait of dense baryonic matter.

In the present paper, an attempt is performed to find out how the phenomenon of condensation of diquark pairs, which in the real 3-color case
corresponds to color superconductivity (CSC) phenomenon, can affect the duality between the CSB and charged PC phases. According to a number of studies
of CSC \cite{alford}, this phase of dense quark matter can be realized in the cores of neutron stars. Naturally, in this case there is an
isospin asymmetry (different densities of $u$ and $d$ quarks) of quark matter. Moreover, it is also %there are ideas that the conditions  it is also 
under the influence of strong magnetic field,
leading to chiral asymmetry of quark medium (see, e.g., the discussion in Ref. \cite{kkz2}). To simplify the consideration of the problem,
in our recent paper \cite{2color} the phase structure of the 2-color and 2-flavor massless NJL model was investigated in the mean-field
approximation at three nonzero chemical potentials, $\mu_B\ne 0$, $\mu_{I}\ne 0$ and $\mu_{I5}\ne 0$. It is well known that at low energies
this model is equivalent to the 2-color QCD (see, e.g., Ref. \cite{weise}) with the same set of chemical potentials, and its simplest particle
excitations are $\sigma$ and $\pi$ mesons and colorless diquark baryons with zero spin. So, in the ground state of the $N_c=2$ system under
consideration there can be a condensation of $\sigma$ particles, and in this case the CSB phase is realized. If $\pi^\pm$ are condensed --
the charged PC phase is observed. Finally, the condensation of baryonic colorless diquarks leads to the phase of quark matter with spontaneous breaking
of baryonic $U(1)_B$ symmetry, and we call it the baryonic superfluid (BSF) phase. It is shown in the paper \cite{2color} that one more, diquark,
channel of quark interactions does not spoil at all the dual symmetry between CSB and charged PC phenomena at large $\mu_B$. Moreover, in this case there appear two additional dual symmetries of the $(\mu_{B},\mu_{I},\mu_{I5})$-phase portrait of the model: between the BSF and CSB phases, as well as between the BSF and charged PC phases.

We emphasize once again that in the paper \cite{2color} the chiral asymmetry of two-color quark matter was taken into account in the form
when only $\mu_{I5}$ is not equal to zero. However, in real systems (heavy-ion collisions, neutron stars) the chiral asymmetry of the medium
in the form when $\mu_5\ne 0$ could play a key role \cite{ruggieri,andrianov,cao,braguta}. In most the scenarios where chiral isospin $\mu_{I5}$ imbalance is prompted, as a rule the chiral $\mu_{5}$ imbalance is also nonzero. Furthermore, in heavy-ion collisions due to large temperatures and non-trivial gluon configurations chiral imbalance $\mu_{5}$ may appear.  Therefore, it would be interesting to clarify the
situation with the dual symmetries of the phase diagram of this system in the most general case, when all four chemical potentials are
taken into account, i.e. at $\mu_B\ne 0$, $\mu_{I}\ne 0$, $\mu_{I5}\ne 0$ and $\mu_5\ne 0$. And clarify if chiral imbalance $\mu_5$ breaks the dualities and how it fits in the duality picture. Moreover, the purpose of the present paper is to investigate the influence of $\mu_5$ on the phenomenon of diquark condensation.  So the present paper is really a continuation of our previous study \cite{2color} of the properties of the 2-color NJL model. And this time we are just considering its phase structure
taking into account  $\mu_5$ in addition to $\mu_B$, $\mu_{I}$ and $\mu_{I5}$.

The main results and the structure of the paper are as follows. In Sec. II the two-color NJL model and its thermodynamic potential are presented in the mean-field approximation. %symmetries are introduced, symmetry breaking patterns and phases that can occupy the phase diagram is 
In Sec. III the thermodynamic potential is calculated and it is shown that the three dualities found in the case $\mu_5=0$ in Ref. \cite{2color} are also valid in the general case at any $\mu_5\neq0$. It turns out that the full 
$(\mu_B,\mu_{I},\mu_{I5},\mu_5)$-phase diagram of the model is interconnected by the dualities and possesses a very high symmetry. Chiral $\mu_5$ is the only chemical potential that keeps this symmetry intact but is not involved in it itself, it only deforms the whole phase diagram. This deformation should respect the high symmetry that puts rather tight constraints on the possible influence of $\mu_5$ on the phase diagram. 

In Sec. IV the phase diagram itself is studied numerically. Sec. IV A contains the discussion of the case when besides $\mu_5\neq0$, one of the basic chemical potentials $\mu_B$, $\mu_{I}$ and $\mu_{I5}$ is nonzero. Already in this case, one could see two interesting features of chiral $\mu_5$ imbalance, its chameleon nature and its property of being universal catalyzer. It is shown that it can catalyze every phenomena in the system, the catalysis of CSB was discussed in details in the literature before (see, e.g., in Ref. \cite{braguta}), but the catalysis/enhancement  of charged PC and diquark condensation by chiral $\mu_5$ imbalance is its new feature.  Chameleon nature signifies that chiral $\mu_5$ chemical potential can take a role of any chemical potential and have the same influence on the phase structure, being universal catalyzer is one of manifestations of chameleon properties, chiral chemical potential can take a role of isospin one and in equal degree catalyze charged PC, or it can mimic baryon one and catalyze diquark condensation. These properties are implications of high symmetry of the phase diagram that is caused by duality properties. 

In Sec. IV B the regime of small or moderate values of the basic chemical potentials $\mu_B$, $\mu_{I}$ and $\mu_{I5}$ is discussed. One should not treat this regime as only small values of these chemical potentials and it covers quite a bit of the phase diagram just excluding the regime when several, two or three, chemical potentials $\mu_B$, $\mu_{I}$ and $\mu_{I5}$ reach high values. 
In this regime the phase diagram is very concise and elegant, each of the basic chemical potential triplet supports only one specific phenomenon 
($\mu_{I5}$ induces chiral symmetry breaking, $\mu_{I}$ -- charged PC and $\mu_B$ leads to diquark condensation), and the largest of the basic chemical potential settles the corresponding phase. In this regime, $\mu_5$ does not break this correspondence and does not play any role in determining the prevailed phase, but its role is the same as in particular case discussed above, i.e. in Sec. IV A, it is universal catalyzer which {\it enhances/catalyzes} all the phenomena picked by other chemical potentials on equal footing.

Sec. IV C contains the consideration of the regime when several, two or three, basic chemical potentials reach rather high values.
In Sec. IV C 1 we study the case when only two of chemical potentials $\mu_B$, $\mu_{I}$ and $\mu_{I5}$ are large and nonzero. In this case $\mu_5$ can cause non-trivial phases to appear, and, depending on the conditions, as a chameleon it could mimic various chemical potentials and {\it trigger} all the possible phases.
Sec. IV C 2 contains the discussion of the most generic case when all basic chemical potentials are nonzero. One could observe a rather complicated and rich phase structure in this case and chiral $\mu_5$ chemical potential is shown to {\it trigger} rather peculiar phases. For example, diquark condensation by taking the role of baryon chemical potential (at zero baryon chemical potential), which is quite unusual. %, this happens due to the ability of chiral $\mu_5$ chemical potential of taking the role of baryon chemical potential by taking the role of baryon one wts. 
Or charged PC at $\mu_I=0$ by mimicking the property of isospin chemical potential, etc. Also the transitory regime is considered, i.e. borderline region between the discussed above regimes. In this case especially rich phase structure is observed and by changing chemical potentials in a rather narrow range a series of first order phase transitions between all the possible phases of the system could be observed.

\section{Two-color (3+1)-dimensional NJL model and its thermodynamic potential}

In order to obtain an effective 4-quark Lagrangian (which is usually called the NJL Lagrangian) that would reproduce the basic low-energy properties of dense quark matter with isospin and chiral asymmetries and formed by $u$ and $d$ two-color quarks, it is necessary first to integrate out the gluon fields in the generating functional of the corresponding QCD theory. Then, replacing the nonperturbative gluon propagator by a $\delta-$function, one arrives at an effective local chiral four-quark interaction Lagrangian
of the form (color current)$\times$(color current) of the NJL type describing low-energy hadron physics. Finally, by performing a Fierz transformation of this interaction term and taking into account only scalar and pseudo-scalar $(\bar q q)$- as well as scalar $(qq)$-interaction channels, one obtains a four-fermionic
model given by the following Lagrangian (in Minkowski space-time notation) \footnote{The most general Fierz transformed four-fermion interaction includes additional vector and axial-vector $(\bar q
q)$ as well as pseudo-scalar, vector and axial-vector-like $(qq)$-interactions. However, these terms are omitted here for simplicity.}
\begin{eqnarray}
L&=&\bar q \Big [i\hat\partial-m_0\Big ]q+H\Big [(\bar qq)^2+(\bar qi\gamma^5\vec\tau q)^2+
\big (\bar qi\gamma^5\sigma_2\tau_2q^c\big )\big (\overline{q^c}i\gamma^5\sigma_2\tau_2 
q\big )\Big]+\bar q {\cal M}\gamma^0 q, \label{1}
\end{eqnarray}
where (here we use the notations $\mu=\mu_B/2$, $\nu=\mu_I/2$ and $\nu_5=\mu_{I5}/2$)
\footnote{Do not be confused, but below for the values of $\mu$, $\nu$ and $\nu_5$ we will usually use the names 
baryon, isospin and chiral isospin chemical potentials, respectively.}
\begin{eqnarray}
{\cal M}= \mu+\nu\tau_3+\nu_5\gamma^5\tau_3+\mu_5\gamma^5. \label{100}
\end{eqnarray}
In (\ref{1}), the quark field $q\equiv q_{i\alpha}$ is a flavor and color doublet as
well as a four-component Dirac spinor, where $i=1,2$ or $u,d$; $\alpha =1,2$. (Latin and Greek indices refer to flavor and color indices, respectively; spinor indices are omitted.) Furthermore,
we use the notations  $\vec \tau\equiv (\tau_{1}, \tau_{2},\tau_3)$ and $\sigma_2$ for usual Pauli matrices acting in the two-dimensional flavor and color spaces, respectively; $\hat\partial\equiv\gamma^\rho \partial_\rho$; $q^c=C\bar q^T$, $\overline{q^c}=q^T C$ are charge-conjugated spinors, and
$C=i\gamma^2\gamma^0$ is the charge conjugation matrix (the symbol $T$ denotes the transposition operation). The Lagrangian (\ref{1}) is invariant with respect to color $SU(2)_c$ and baryon $U(1)_B$ symmetries. Moreover, in the chiral limit, $m_0=0$, and at zero values of all chemical potentials it has the same Pauli-Gursey flavor $SU(4)$ symmetry as the corresponding two-color QCD.

The Lagrangian $L$ of Eq. (\ref{1}) contains baryon $\mu_B$, isospin $\mu_I$, chiral isospin $\mu_{I5}$, and chiral $\mu_{5}$ chemical potentials.
In other words, this model is able to describe the properties of quark matter with nonzero baryon $n_B=(n_{u}+n_{d})/2\equiv n/2$,
isospin $n_I=(n_{u}-n_{d})/2$, chiral isospin $n_{I5}=(n_{u5}-n_{d5})/2$ and chiral $n_{5}=n_{R}-n_{L}$ densities which are the quantities,
thermodynamically conjugated to chemical potentials $\mu_B$, $\mu_I$, $\mu_{I5}$ and $\mu_{5}$, respectively.
(Here we use the notations $n_f$ and $n_{fL(R)}$ for density of quarks as well as density of 
left(right)-handed quarks, $q_{L/R}=\frac{1\mp\gamma^5}2 q$, with individual
flavor $f=u,d$, respectively. Moreover, $n_{f5}=n_{fR}-n_{fL}$ and $n_{R(L)}=n_{uR(L)}+n_{dR(L)}$. It is also supposed throughout the paper
that quark fields in Eq. (\ref{1}) have a baryon charge equal to 1/2.) Below we need the expressions for the chemical 
potentials $\mu_{fL(R)}$, i.e. quantities which are thermodynamically conjugated to the particle number 
densities $n_{fL(R)}$ of left-handed (right-handed) $f=u,d$ quarks, respectively,
\begin{eqnarray}
\mu_{uL}&=&\mu+\nu+\mu_5+\nu_5,~~\mu_{uR}=\mu+\nu-\mu_5-\nu_5,\nonumber\\
\mu_{dL}&=&\mu-\nu+\mu_5-\nu_5,~~\mu_{dR}=\mu-\nu-\mu_5+\nu_5. \label{1000}
\end{eqnarray}
The quantities (\ref{1000}) can be obtained from Eqs. (\ref{1}) and (\ref{100}). They define Fermi energies for $u_{L(R)}$ and $d_{L(R)}$ quarks. Note that at $m_0=0$ the Lagrangian
(\ref{1}) is no longer invariant with respect to Pauli-Gursey $SU(4)$ symmetry. Due to the terms with chemical potentials, this symmetry
is reduced to the abelian $U(1)_B$, $U(1)_{I_3}$ and $U(1)_{AI_3}$
groups, where
\begin{eqnarray}
U(1)_B:~q\to\exp (\mathrm{i}\alpha/2) q;~
U(1)_{I_3}:~q\to\exp (\mathrm{i}\alpha\tau_3/2) q;~
U(1)_{AI_3}:~q\to\exp (\mathrm{i}
\alpha\gamma^5\tau_3/2) q.
\label{2001}
\end{eqnarray}
Moreover, the quantities $n_B$, $n_I$ and $n_{I5}$ are the ground state expectation values of the densities of conserved charges corresponding
to $U(1)_B$, $U(1)_{I_3}$ and $U(1)_{AI_3}$ symmetry groups. So we have from (\ref{2001}) that
$n_B=\vev{\bar q\gamma^0q}/2$, $n_I=\vev{\bar q\gamma^0\tau^3 q}/2$ and $n_{I5}=\vev{\bar q\gamma^0\gamma^5\tau^3 q}/2$.
However, the chiral chemical potential $\mu_5$ does not correspond to a conserved quantity of the model (\ref{1}). It is usually introduced in order to describe a system on the time scales when all chirality changing processes are finished in the system, so it is in the state of thermodynamical equilibrium with some fixed value of the chiral density $n_5$ \cite{andrianov}. The ground state expectation values of $n_B$,
$n_I$, $n_{I5}$ and $n_{5}$ can be found by differentiating the thermodynamic potential (TDP) of the system (\ref{1}) with respect to the corresponding chemical potentials. The goal of the present paper is the investigation of the ground state properties (or phase structure) of the NJL model (\ref{1}) and its dependence on the chemical potentials  $\mu_B$, $\mu_I$, $\mu_{I5}$ and $\mu_{5}$.

To find the TDP, we starting from a semibosonized (linearized) version of the Lagrangian (\ref{1}) that contains auxiliary bosonic fields $\sigma (x)$, $\vec\pi  =(\pi_1 (x),\pi_2 (x),\pi_3 (x))$, $\Delta (x)$ and $\Delta^* (x)$ and has the following form
\begin{eqnarray}
\widetilde L=\bar q \Big [i\hat\partial-m_0+{\cal M}\gamma^0 -\sigma -i\gamma^5\vec\tau\vec\pi\Big ]q-\frac{\sigma^2+\vec\pi^2+
\Delta^*\Delta}{4H}-\frac{\Delta}{2}\Big [\bar qi\gamma^5\sigma_2\tau_2q^c\Big ]-
\frac{\Delta^*}{2}\Big [\overline{q^c}i\gamma^5\sigma_2\tau_2 q\Big ], \label{2}
\end{eqnarray}
where ${\cal M}$ is presented in Eq. (\ref{100}). Clearly, the Lagrangians (\ref{1}) and (\ref{2}) are equivalent, as can be seen by using the Euler-Lagrange equations of motion for bosonic fields which take the form
\begin{eqnarray}
\sigma (x)=-2H(\bar qq),&~&\Delta (x)= -2H\Big [\overline{q^c}i\gamma^5\sigma_2\tau_2 q\Big ]=-2H\Big [q^TCi\gamma^5\sigma_2\tau_2 q\Big ],\nonumber\\
~\vec\pi(x)=-2H(\bar qi\gamma^5\vec\tau q),&~&
\Delta^*(x)=-2H\Big [\bar qi\gamma^5\sigma_2\tau_2q^c\Big ]=-2H\Big [\bar qi\gamma^5\sigma_2\tau_2C\bar q^T\Big ].
\label{3}
\end{eqnarray}
It is easy to see from Eq. (\ref{3}) that $\sigma(x)$ and $\pi_a(x)$ ($a=1,2,3$) are Hermitian, i.e. real, bosonic fields, whereas $\Delta^*(x)$ and $\Delta(x)$ are Hermitian conjugated to each other. Indeed, one can check that $(\sigma(x))^\dagger=\sigma(x)$,
$(\pi_a(x))^\dagger=\pi_a(x)$, $(\Delta(x))^\dagger=\Delta^*(x)$ and $(\Delta^*(x))^\dagger=\Delta(x)$, where the superscript symbol $\dagger$ denotes the Hermitian conjugation. Note that the composite bosonic field $\pi_3 (x)$ can be identified with the physical $\pi^0(x)$-meson field, whereas the physical $\pi^\pm (x)$-meson fields are the following combinations of the composite fields, $\pi^\pm (x)=(\pi_1 (x)\mp i\pi_2 (x))/\sqrt{2}$. It is clear that the groung state expectation values of all boson fields (\ref{3})
are $SU(2)_c$ invariants, hence in this model the color symmetry can not be broken dynamically. If the ground state expectation values $\vev{\sigma (x)}\ne 0$ or $\vev{\pi_0 (x)}\ne 0$, then chiral symmetry $U(1)_{AI_3}$ of the model (\ref{1}) is broken spontaneously. If in the ground state we have $\vev{\pi_{1,2} (x)}\ne 0$, then isospin  $U(1)_{I_3}$ is broken spontaneously. This phase of quark matter is called the charged pion condensation (PC) phase. Finally, if $\vev{\Delta (x)}\ne 0$, then in the system spontaneous breaking of the baryon $U(1)_B$ symmetry occurs, and the baryon superfluid (BSF) phase is realized in the model.
%Note that in the BSF phase the electromagnetic $U(1)_Q$ symmetry remains intact.

Introducing the Nambu-Gorkov bispinor field $\Psi$, where
\begin{equation}
\Psi=\left({q\atop q^c}\right),~~\Psi^T=(q^T,\bar q C^{-1});~~
\quad \overline\Psi=(\bar q,\overline{q^c})=(\bar q,q^T C)=\Psi^T \left
(\begin{array}{cc}
0~~,&  C\\
C~~, &0
\end{array}\right )\equiv\Psi^T Y,
\label{5}
\end{equation}
one can bring the auxiliary Lagrangian (\ref{2}) to the following form
\begin{eqnarray}
\widetilde L=-\frac{\sigma^2+\vec\pi^2+\Delta^*\Delta}{4H}+\frac 12\Psi^T(YZ)\Psi, \label{12}
\end{eqnarray}
where matrix $Y$ is given in Eq. (\ref{5}) and
\begin{equation}
Z=\left (\begin{array}{cc}
D^+, & K\\
K^*~~, &D^-
\end{array}\right )\equiv \left (\begin{array}{cc}
i\hat\partial-m_0+{\cal M}\gamma^0 -\sigma -i\gamma^5\vec\tau\vec\pi, & -i\gamma^5\sigma_2\tau_2\Delta\\
~~~~~~~~-i\gamma^5\sigma_2\tau_2\Delta^*~~~~~~~~~~~~, &i\hat\partial-m_0-\gamma^0{\cal M} -\sigma -i\gamma^5(\vec\tau)^T\vec\pi
\end{array}\right ).\label{13}
\end{equation}
Notice that matrix elements of the 2$\times$2 matrix $Z$, i.e. the quantities $D^\pm$, $K$ and $K^*$, are the nontrivial operators in the (3+1)-dimensional coordinate, four-dimensional spinor, 2-dimensional flavor and ($N_c=2$)-dimensional color spaces. Then, in the one fermion-loop (or mean-field) approximation, the effective action ${\cal S}_{\rm {eff}}(\sigma,\vec\pi,\Delta,\Delta^{*})$ of the model (\ref{1})-(\ref{2}) (this quantity is the generating functional of a one-particle irreducible Green functions of boson fields (\ref{3})) is expressed by means of the path integral over quark fields:
\begin{eqnarray}
\exp(i {\cal S}_{\rm {eff}}(\sigma,\vec\pi,\Delta,\Delta^{*}))=
  N'\int[d\bar q][dq]\exp\Bigl(i\int\widetilde L\,d^4 x\Bigr),\label{14}
\end{eqnarray}
where $N'$ is a normalization constant and
\begin{eqnarray}
&&{\cal S}_{\rm {eff}}
(\sigma,\vec\pi,\Delta,\Delta^{*})
=-\int d^4x\left [\frac{\sigma^2(x)+\vec\pi^2(x)+|\Delta(x)|^2}{4H}\right ]+
\widetilde {\cal S}_{\rm {eff}}.
\label{15}
\end{eqnarray}
The quark contribution to the effective action, i.~e.\  the term
$\widetilde {\cal S}_{\rm {eff}}$ in (\ref{15}), is given by:
\begin{equation}
\exp(i\tilde {\cal S}_{\rm {eff}})=N'\int [d\bar
q][dq]\exp\Bigl(\frac{i}{2}\int\Big [\Psi^T(YZ)\Psi\Big ]d^4 x\Bigr).
\label{16}
\end{equation}
Note that in Eqs. (\ref{14})-(\ref{16}) we have used the expression (\ref{12}) for the auxiliary Lagrangian $\widetilde L$.
Since the integration measure in Eq. (\ref{16}) obeys the relation $[d\bar q][dq]=$ $[d q^c][dq]=$ $[d\Psi]$, we have from it
\begin{equation}
\exp(i\tilde {\cal S}_{\rm {eff}})=
  \int[d\Psi]\exp\left\{\frac i2\int\Psi^T(YZ)\Psi
  d^4x\right\}=\mbox {det}^{1/2}(YZ)=\mbox {det}^{1/2}(Z),\label{17}
\end{equation}
where the last equality is valid due to the evident relation $\det
Y=1$. Then, using the Eqs (\ref{15}) and (\ref{17}) one can obtain the following
expression for the effective action (\ref{15}):
\begin{equation}
{\cal S}_{\rm
{eff}}(\sigma,\vec\pi,\Delta,\Delta^{*})
=-\int d^4x\left[\frac{\sigma^2(x)+\vec\pi^2(x)+|\Delta(x)|^2}{4H}\right]-\frac i2\ln\mbox {det}(Z).
%{\rm Tr}_{sfcx\scriptstyle NG}\ln (Z+W).
\label{18}
\end{equation}
Starting from Eq. (\ref{18}), one can define in the
mean-field approximation the thermodynamic potential (TDP) $\Omega(\sigma,\vec\pi, \Delta, \Delta^{*})$
of the model (\ref{1})-(\ref{2}),
\begin{equation}
{\cal S}_{\rm {eff}}~\bigg
|_{~\sigma,\vec\pi,\Delta,\Delta^{*}=\rm {const}}
=-\Omega(\sigma,\vec\pi,\Delta,\Delta^{*})\int d^4x.
\label{19}
\end{equation}
The ground state expectation values (mean values) of the fields:
$\vev{\sigma(x)}\equiv\sigma,~\vev{\vec\pi(x)} \equiv\vec\pi,~\vev {\Delta(x)}\equiv\Delta,~
\vev{\Delta^{*}(x)}\equiv\Delta$, are the solutions of the gap equations for the TDP $\Omega$ (below, in our approach all ground state expectation values $\sigma,\vec\pi,\Delta,\Delta^*$ do not depend on coordinates $x$):
\begin{eqnarray}
\frac{\partial\Omega}{\partial\pi_a}=0,~~~~~
\frac{\partial\Omega}{\partial\sigma}=0,~~~~~
\frac{\partial\Omega}{\partial\Delta}=0,~~~~~
\frac{\partial\Omega}{\partial\Delta^{*}}=0.
\label{20}
\end{eqnarray}
Since the matrix $Z$ in Eq. (\ref{18}) has an evident 2$\times$2 block structure (see in Eq. (\ref{13})), one can use there a general formula
\begin{eqnarray}
\det\left
(\begin{array}{cc}
A~, & B\\
C~, & D
\end{array}\right )=\det [-CB+CAC^{-1}D]=\det
[DA-DBD^{-1}C],\label{21}
\end{eqnarray}
and find that (taking into account the relation $\tau_2\vec\tau\tau_2=-\vec\tau^T$ and assuming that all bosonic fields do not depend on $x$)
\begin{eqnarray}
&&\det(Z)\equiv\det\left
(\begin{array}{cc}
D^+~, & K\\
K^*~, & D^-
\end{array}\right )=\det\big(-K^*K+K^*D^+K^{*-1}D^-\big)\nonumber\\
&&=\det \Big [\Delta^*\Delta+\big (-i\hat\partial-m_0-\widetilde{\cal M}\gamma^0 -\sigma +i\gamma^5(\vec\tau)^T\vec\pi\big )\big (i\hat\partial-m_0-
\gamma^0{\cal M} -\sigma -i\gamma^5(\vec\tau)^T\vec\pi\big )\Big ],\label{22}
\end{eqnarray}
where
\begin{eqnarray}
\widetilde{\cal M}=\mu+\mu_5\gamma^5-\nu\tau_3-\nu_{5}\gamma^5\tau_3. \label{23}
\end{eqnarray}
Obviously, the quantity which is in the square brackets of Eq. (\ref{22}) is proportional to the unit operator in the $N_c$-color space. (Below, in all numerical calculations we put $N_c=2$.) Hence,
\begin{eqnarray}
&&\det(Z)={\rm det}^{N_c}{\cal D}\equiv{\rm det}^{N_c}\left
(\begin{array}{cc}
D_{11}~, & D_{12}\\
D_{21}~, & D_{22}
\end{array}\right ),\label{24}
\end{eqnarray}
where ${\cal D}$ is the $2\times 2$ matrix in the 2-dimensional flavor space (its matrix elements $D_{kl}$ are the nontrivial operators in the 4-dimensional spinor and in the (3+1)-dimensional coordinate spaces). Using this expression for $\det(Z)$ in Eq. (\ref{18}) when $\sigma,\vec\pi,\Delta,\Delta^*$ do not depend on coordinates $x$, and taking into account the well-known technique for calculating determinants of operators (see, for example, Eq. (A6) of Appendix A from Ref. \cite{2color}), we find
\begin{eqnarray}
{\cal S}_{\rm
{eff}}(\sigma,\pi_a,\Delta,\Delta^{*})~\bigg
|_{~\sigma,\vec\pi,\Delta,\Delta^{*}=\rm {const}}
&=&-\frac{\sigma^2+\vec\pi^2+|\Delta|^2}{4H}\int d^4x-\frac {iN_c}2\ln\mbox {det}{\cal D}\nonumber\\
=-\frac{\sigma^2+\vec\pi^2+|\Delta|^2}{4H}\int d^4x&-&\frac {iN_c}2\int\frac{d^4p}{(2\pi)^4}\ln\det\overline{\cal D}(p)\int d^4x,
%\mbox {det}.
%{\rm Tr}_{sfcx\scriptstyle NG}\ln (Z+W).
\label{32}
\end{eqnarray}
where the 2$\times$2 matrix $\overline{\cal D}(p)$ is the momentum space representation of the matrix ${\cal D}$ of Eq. (\ref{24}). Its matrix elements $\overline{\cal D}_{kl}(p)$ have the following form
\begin{eqnarray}
\overline{D}_{11}(p)&=&|\Delta |^2-p^2+(\mu+\mu_5\gamma^5)\big[\hat p \gamma^0-\gamma^0\hat p \big ]+
(\nu+\nu_5\gamma^5)\big[\hat p \gamma^0+\gamma^0\hat p\big ]+\vec\pi^2+M^2\nonumber\\
&+&2M\gamma^0(\mu+\nu_5\gamma^5)+(\mu+\mu_5\gamma^5)^2-(\nu+\nu_5\gamma^5)^2
+2i\mu\gamma^0\gamma^5\pi_3+2i\nu_5\gamma^0\pi_3,\nonumber\\
\overline{D}_{22}(p)&=&|\Delta |^2-p^2+(\mu+\mu_5\gamma^5)\big[\hat p \gamma^0-\gamma^0\hat p \big ]
-(\nu+\nu_5\gamma^5)\big[\hat p\gamma^0+\gamma^0\hat p\big ]+\vec\pi^2+M^2\nonumber\\
&+&2M\gamma^0(\mu-\nu_5\gamma^5)+(\mu+\mu_5\gamma^5)^2-(\nu+\nu_5\gamma^5)^2
-2i\mu\gamma^0\gamma^5\pi_3+2i\nu_5\gamma^0\pi_3,\nonumber\\
\overline{D}_{12}(p)&=&2i\mu\gamma^0\gamma^5\big(\pi_1+i\pi_2\big)+2\nu\gamma^0\gamma^5\big(\pi_2-i\pi_1\big)=
2\gamma^0\gamma^5(\nu-\mu)[\pi_2-i\pi_1],\nonumber\\
\overline{D}_{21}(p)&=&2i\mu\gamma^0\gamma^5\big(\pi_1-i\pi_2\big)+2\nu\gamma^0\gamma^5\big(i\pi_1+\pi_2\big)=
2\gamma^0\gamma^5(\nu+\mu)[\pi_2+i\pi_1],
\label{33}
\end{eqnarray}
where $M\equiv m_0+\sigma$, $p^2=p^\rho p_\rho$, $\hat p=\gamma^\rho p_\rho$. Using in Eq. (\ref{32}) again the general relation (\ref{21}), we have
\begin{eqnarray}
&&\det\overline{D}(p)\equiv {\rm det}\left
(\begin{array}{cc}
\overline{D}_{11}(p)~, & \overline{D}_{12}(p)\\
\overline{D}_{21}(p)~, & \overline{D}_{22}(p)
\end{array}\right )\nonumber\\
&=&\det\Big[-\overline{D}_{21}(p)\overline{D}_{12}(p)+\overline{D}_{21}(p)\overline{D}_{11}(p)\big(\overline{D}_{21}(p)\big)^{-1}
\overline{D}_{22}(p)\Big]\equiv\det L(p).\label{34}
\end{eqnarray}
Notice that the matrix $L(p)$, i.e. the expression in square brackets of Eq. (\ref{34}), is indeed a 4$\times$4 matrix in 4-dimensional spinor space only, which is composed of 4$\times$4 matrices $\overline{D}_{ij}(p)$ (see in Eq. (\ref{33})). Now, taking into account the definition (\ref{19}) and using the Eqs. (\ref{32})-(\ref{34}), it is possible to obtain in the mean-field approximation the TDP of the model,
\begin{eqnarray}
\Omega(M,\vec\pi,\Delta,\Delta^{*})
&=&\frac{(M-m_0)^2+\vec\pi^2+|\Delta|^2}{4H}+\frac {iN_c}2\int\frac{d^4p}{(2\pi)^4}\ln\det L(p).
%\mbox {det}.
%{\rm Tr}_{sfcx\scriptstyle NG}\ln (Z+W).
\label{35}
\end{eqnarray}
Since $\det L(p)=\lambda_1(p)\lambda_2(p)\lambda_3(p)\lambda_4(p)$, where $\lambda_i(p)$ ($i=1,...,4$) are four eigenvalues of the 4$\times$4 matrix $L(p)$, in the following, in order to find the TDP of the model in various cases, we will first of all find the eigenvalues of the matrix $L(p)$. Then, after integration in Eq. (\ref{35}) over $p_0$, this TDP is used in some numerical calculations with sharp three-momentum cutoff $\Lambda=657$ MeV (i.e. it is assumed below that the integration over three-momentum $\vec p$ occurs over the region $|\vec p|<\Lambda$) at $H=7.23$ GeV$^{-2}$ and $m_0=5.4$ MeV \cite{brauner1,andersen2}. Moreover, we study also the phase structure of the model in the chiral limit, $m_0=0$, at the same values of $\Lambda$ and $H$.

Note that at first glance, the TDP (\ref{35}) looks like a function of six variables (condensates), $M$, $\vec\pi$, $\Delta$ and
$\Delta^{*}$. But due to a symmetry of the model, the number of condensates that characterize the ground state of a system may be reduced.
Indeed, at $m_0=0$ and zero chemical potentials the Lagrangian (\ref{1}) is invariant under $SU(4)\times U(1)_B\times SU(2)_c$ group. As a
consequence of this symmetry, the TDP is a function of only one single variable $(M^2+|\Delta|^2+\vec\pi^2)$. And this fact significantly
simplifies the analysis of the function (\ref{35}) on the global minimum. If nonzero chemical potentials are taken into consideration, then
in the chiral limit the symmetry of the Lagrangian (\ref{1}) reduces to $U(1)_B\times U(1)_{I_3}\times U(1)_{AI_3}$ (plus color $SU(2)_c$,
which in our consideration is not violated at all). As a result, we see that at $m_0=0$ and nonzero chemical potentials the TDP of the model
depends only on the $|\Delta|^2$, $\pi_1^2+\pi_2^2$ and $M^2+\pi_0^2$ field combinations, correspondingly. So without loss of generality of
consideration, in the chiral limit we can put $\pi_0=0$ and $\pi_2=0$. However, at $m_0\ne 0$ and at nonzero chemical potentials the symmetry
of the model Lagrangian reduces to $U_B(1)\times U_{I_3}(1)$, i.e. in this case the TDP depends on $|\Delta|$, $M,\pi_0$ and $\pi_1^2+\pi_2^2$.
So in this case without loss of generality we can also put $\pi_2=0$. Moreover, at $m_0\ne 0$, as it was argued in our previous paper
\cite{2color}, it is possible to put $\pi_0=0$ as well. Hence, below throughout the paper we suppose that the TDP (\ref{35}) is a
function of only $M,\pi_1$ and $|\Delta|$ condensates. Others are zero.

\section{Calculation of the TDP (\ref{35}) and its duality properties.}
\subsection{The case of $\mu\ne 0$, $\nu\ne 0$, $\nu_5\ne 0$, but $\mu_5=0$}

First, let us suppose that chiral chemical potential $\mu_5$ is equal to zero. The rest of chemical potentials, i.e. $\mu$, $\nu$ and $\nu_5$, we will call the basic chemical potentials, since (i) they correspond to the conserved charges of the model, and (ii) it is they that largely determine the phase structure of the model. %(As it will be shown below, the $\mu_5\ne 0$ only catalyses/enhances the phenomena realized in the model at $\mu_5=0$.)
In this case when $\mu\ne 0$, $\nu\ne 0$ and $\nu_5\ne 0$ the matrix $L(p)$ of Eqs. (\ref{34}) and (\ref{35}) has four different
eigenvalues $\lambda_{i}(p)$ (they can be found with the help of any program of analytical calculations),
\begin{eqnarray}
&&\lambda_{1,2}(p)=N_1\pm 4\sqrt{K_1},~~\lambda_{3,4}(p)=N_2\pm 4\sqrt{K_2},
\label{56}
\end{eqnarray}
where
\begin{eqnarray}
\hspace{-1cm}N_2=N_1+16\mu\nu\nu_5|\vec p|,~~K_2=K_1+8\mu\nu\nu_5|\vec p|p_0^4-8\mu\nu\nu_5|\vec p|p_0^2\big (M^2+\pi_1^2+|\Delta|^2
+|\vec p|^2+\mu^2+\nu^2-\nu_5^2\big ),&&
\label{57}
\end{eqnarray}
\begin{eqnarray}
%\hspace{-1cm}
K_1=\nu_5^2p_0^6-p_0^4\Big [2\nu_5^2 \big(|\Delta|^2+\pi_1^2+M^2+|\vec p|^2+\nu^2+\mu^2-\nu_5^2\big)+4\mu\nu\nu_5|\vec p|\Big ]+
p_0^2 \Big\{
\nu_5^6+2\nu_5^4 \big (M^2-|\Delta|^2-\pi_1^2&&~~~~~~~~~~~~~~\nonumber\\
-\nu^2-\mu^2-|\vec p|^2\big )+4\mu^2\nu^2\big(M^2+
|\vec p|^2\big)+4|\vec p|\mu\nu\nu_5\big(|\Delta|^2+\pi_1^2+M^2+|\vec p|^2+\nu^2+\mu^2-\nu_5^2\big)&&\nonumber\\
+\nu_5^2\Big [\big(|\Delta|^2+\pi_1^2+|\vec p|^2+\nu^2+\mu^2\big)^2+2|\vec p|^2M^2+M^4+2M^2\big(|\Delta|^2-\nu^2+\pi_1^2-\mu^2\big)\Big ]\Big\},&&
\label{58}
\end{eqnarray}
\begin{eqnarray}
N_1&=&p_0^4-2p_0^2\Big [|\Delta|^2+\pi_1^2+M^2+|\vec p|^2+\nu^2+\mu^2-3\nu_5^2\Big ]+
\nu_5^4-2\nu_5^2\Big [|\Delta|^2+\pi_1^2+|\vec p|^2+\nu^2+\mu^2-M^2\Big ]\nonumber\\
&-&8\mu\nu\nu_5|\vec p|+\left (|\vec p|^2+M^2+\pi_1^2+|\Delta|^2-\mu^2-\nu^2\right )^2-
4\left (\mu^2\nu^2-\pi_1^2\nu^2-|\Delta|^2\mu^2\right ),
\label{59}
\end{eqnarray}
and the TDP (\ref{35}) has the form
\begin{eqnarray}
\hspace{-1.5cm} \Omega(M,\pi_1,|\Delta|)
&=&\frac{(M-m_0)^2+\pi_1^2+|\Delta|^2}{4H}+\frac {iN_c}2\int\frac{d^4p}{(2\pi)^4}\Big\{\ln\Big(\lambda_1(p)\lambda_2(p)
\Big)+\ln\Big(\lambda_3(p)\lambda_4(p)\Big )\Big\}.
\label{73}
\end{eqnarray}
It is possible to show that the TDP (\ref{73}) is an even with respect to each of the three transformations, $\mu\to -\mu$, $\nu\to -\nu$ and
$\nu_5\to -\nu_5$. Indeed, if, e.g., $\mu\to -\mu$ then, as it follows from Eqs. (\ref{56})-(\ref{59}), we have $\lambda_1(p)\leftrightarrow
\lambda_3(p)$
and $\lambda_2(p)\leftrightarrow\lambda_4(p)$. Hence, when the sign of the chemical potential $\mu$ changes, the TDP (\ref{73})
itself remains invariant, etc. It means that without loss of generality we can use only the positive values of the chemical potentials.
More interesting is the fact that each of the eigenvalues $\lambda_i(p)$ (\ref{56}) is invariant with respect to the so-called dual transformation ${\cal D}_1$,
\begin{eqnarray}
{\cal D}_1: ~~~~\mu\longleftrightarrow\nu,~~~\pi_1\longleftrightarrow |\Delta|.
\label{55}
\end{eqnarray}
As a result, at $\mu_5=0$, and even at $m_0\ne 0$, the whole TDP (\ref{35}) or (\ref{73}) of the model is also invariant under the transformation (\ref{55}).
Moreover, using any program of analytical calculations, it is possible to establish that each of the products $\lambda_1(p)\lambda_2(p)$ and $\lambda_3(p)\lambda_4(p)$ is invariant in addition with respect to the following two dual discrete transformations ${\cal D}_2$ and ${\cal D}_3$, where
\begin{eqnarray}
{\cal D}_2: ~~\mu\longleftrightarrow\nu_5,~~M\longleftrightarrow |\Delta|;~~~~~~{\cal D}_3:
~~\nu\longleftrightarrow\nu_5,~~M\longleftrightarrow \pi_1.
\label{60}
\end{eqnarray}
This fact was proved in Ref. \cite{2color}. As a result, we see that at $m_0=0$ the TDP (\ref{73}) is invariant under the dual transformations ${\cal D}_2$ and ${\cal D}_3$ (\ref{60}) in addition to ${\cal D}_1$ (\ref{55}). (Note that at $m_0\ne 0$ it is invariant only under the ${\cal D}_1$ (\ref{55}) transformation.)

\subsection{The case of all nonzero chemical potentials}

Now, let us consider the case when $\mu_5\ne 0$ is taken into account in addition to the basic
chemical potentials $\mu\ne 0$, $\nu\ne 0$ and $\nu_5\ne 0$.
It is possible to show that in this case the matrix $L(p)$ of Eqs. (\ref{34}) and (\ref{35}) has four different eigenvalues $\widetilde\lambda_{i}(p)$,
\begin{eqnarray}
&&\widetilde\lambda_{1,2}(p)=\lambda_{1,2}(p)\Big |_{|\vec p|\to|\vec p|-\mu_5},~~\widetilde\lambda_{3,4}(p)=
\lambda_{3,4}(p)\Big |_{|\vec p|\to|\vec p|+\mu_5},
\label{71}
\end{eqnarray}
where $\lambda_i(p)$ are the eigenvalues (\ref{56}) of the $L(p)$ at $\mu_5=0$. Hence in the most general case we obtain the following expression for the TDP (\ref{35})
\begin{eqnarray}
\hspace{-1.5cm} \Omega(M,\pi_1,|\Delta|)
&=&\frac{(M-m_0)^2+\pi_1^2+
|\Delta|^2}{4H}+\frac {iN_c}2\int\frac{d^4p}{(2\pi)^4}\Big\{\ln\Big(\widetilde\lambda_1(p)\widetilde\lambda_2(p)
\Big)+\ln\Big(\widetilde\lambda_3(p)\widetilde\lambda_4(p)\Big )\Big\}.
\label{72}
\end{eqnarray}
Due to the relations (\ref{71}), it is clear that the chemical potential $\mu_5$ does not spoil
the dual symmetries inherent to the TDP (\ref{73}) in
the case $\mu_5=0$. So, at $m_0\ne 0$ the TDP (\ref{35}) or (\ref{72}) is invariant with respect to the dual ${\cal D}_1$ (\ref{55})
transformation, whereas in the chiral limit, $m_0=0$, it is invariant with respect to the
${\cal D}_2$ and ${\cal D}_3$ (\ref{60}) duality transformations, in addition. It is one of the main
results of the paper.

It is clear directly from the relations (\ref{71}) that if one of the chemical potentials is equal to zero, then $\Omega (M,\pi_1,|\Delta|)$
is an even function with respect to each of the rest nonzero chemical potentials.
%\footnote{For example, if $\mu_5=0$, then the TDP (\ref{9}) is symmetric under the transformation $\mu\to-\mu$.
%Indeed, in this case simultaneously with $\mu\to-\mu$ one should perform in the integral (\ref{9}) the $p_0\to-p_0$ and $p_1\to-p_1$
%change of variables. As a result, one can easily see that the expression (\ref{9}) remains intact. If $\nu\to-\nu$ at $\mu_5=0$,
%then it is enough to change $p_1\to-p_1$ in the integral (\ref{9}) in order to provide the invariance of this integral, etc.}
However, if all four chemical potentials $\mu$, $\nu$, $\nu_5$, and $\mu_5$ are nonzero, then it is easily seen from relations (\ref{71})
and (\ref{56})-(\ref{59}) that the TDP (\ref{72}) is invariant with respect to each of the following six transformations, in each of them two
chemical potentials change their sign simultaneously: (i) $\{\nu\to-\nu;~\nu_5\to-\nu_5\}$, (ii) $\{\nu\to-\nu;~\mu_5\to-\mu_5\}$, (iii)
$\{\nu_5\to-\nu_5;~\mu_5\to-\mu_5\}$, (iv) $\{\mu\to-\mu;~\mu_5\to-\mu_5\}$, (v) $\{\mu\to-\mu;~\nu\to-\nu\}$, and (vi)
$\{\mu\to-\mu;~\nu_5\to-\nu_5\}$. The invariance of the TDP (\ref{72}) under the transformations (i)--(vi) can help to
simplify the analysis of the phase portrait of the model. In particular, it is sufficient to study the phase structure of the model
only, e.g.,  in the case when arbitrary three of the four chemical potentials have positive signs, whereas the sign of the rest chemical
potential is not fixed. Then, applying to a phase diagram with this particular distribution of the chemical potential signs one or several
transformations (i)--(vi), it is possible to find a phase portrait of the model at an arbitrary distribution of chemical potential signs.
Hence, in the following we will study the phase diagram of the model only at $\mu\ge 0$, $\nu\ge 0$, $\nu_5\ge 0$ and for arbitrary sign of $\mu_5$.

Note that in our previous article \cite{2color}, where the chiral asymmetry of the two-color dense quark system was investigated in the form of only $\mu_{I5}\ne 0$ but $\mu_5=0$, it was argued that in the chiral limit, $m_0=0$, for sufficiently low values of the chemical potentials (say at $\mu,\nu,\nu_5 <1$ GeV) at the global minimum point (GMP) $(M,\pi_1,|\Delta|)$ of the TDP (\ref{73}), there can be no more than one
nonzero coordinates, i.e. condensates or order parameters. (The particular  argument could be that in previous investigations there have not been found
(mixed) phases with several nonzero condensates, e.g., in the three-color NJL model \cite{kkz2} and in the two-color one \cite{son2,andersen2}.) Therefore, with such a restriction on chemical potentials, in the chiral limit only four different phases can be realized in the system.
(I) If GMP has the form $(M\ne 0,\pi_1=0,|\Delta|=0)$, then the chiral symmetry breaking (CSB) phase appears in the model. (II) If it has
the form $(M=0,\pi_1\ne 0,|\Delta|=0)$, the charged pion condensation (PC) phase is realized. (III) When the GMP looks like
$(M=0,\pi_1=0,|\Delta|\ne 0)$, it corresponds to the baryon superfluid (BSF) or diquark condensation phase. And finally, (IV) the GMP of the
form $(M=0,\pi_1=0,|\Delta|=0)$ corresponds to a symmetrical phase with all zero condensates.

In a similar way, in the present paper we suppose that if, in addition, the chemical potential $\mu_5$ is taken into account, then the GMP of the TDP (\ref{72}) has the same structure, and only four above mentioned phases (I)-(IV) are allowed to exist in the system.
Thanks to this structure of the global minimum point, we see that in the region of relatively
low values of the chemical potentials it is enough to study not the whole TDP (\ref{72}), but only its projections on the condensate axis,
$F_1(M)\equiv\Omega(M,\pi_1=0,|\Delta|=0)$, $F_2(\pi_1)\equiv\Omega(M=0,\pi_1,|\Delta|=0)$ and
$F_3(|\Delta|)\equiv\Omega(M=0,\pi_1=0,|\Delta|)$, where
\begin{eqnarray}
%\hspace{-1cm}
F_1(M)=\frac{M^2}{4H}
&-&\frac{N_c}2\sum_{\pm}\int\frac{d^3p}{(2\pi)^3}\Big [\big |\mu+\nu\pm\sqrt{M^2+(|\vec p|-\mu_5-\nu_5)^2}\big|
+\big |\mu-\nu\pm\sqrt{M^2+(|\vec p|-\mu_5+\nu_5)^2}\big|\nonumber\\&&%\hspace{-1cm}
%+\big |\mu-\nu\pm\sqrt{M^2+(|\vec p|-\nu_5)^2}\big|
+\big |\mu+\nu\pm\sqrt{M^2+(|\vec p|+\mu_5+\nu_5)^2}\big|+
\big |\mu-\nu\pm\sqrt{M^2+(|\vec p|+\mu_5-\nu_5)^2}\big|\Big ],
\label{1890}
\end{eqnarray}
\begin{eqnarray}
F_2(\pi_1)
=\frac{\pi_1^2}{4H}
&-&\frac{N_c}2\sum_{\pm}\int\frac{d^3p}{(2\pi)^3}\Big [\big |\mu+\nu_5\pm\sqrt{\pi_1^2+(|\vec p|-\mu_5-\nu)^2}\big|
+\big |\mu-\nu_5\pm\sqrt{\pi_1^2+(|\vec p|-\mu_5+\nu)^2}\big|\nonumber\\&&
+\big |\mu+\nu_5\pm\sqrt{\pi_1^2+(|\vec p|+\mu_5+\nu)^2}\big|+
\big |\mu-\nu_5\pm\sqrt{\pi_1^2+(|\vec p|+\mu_5-\nu)^2}\big|\Big ],
\label{1900}
\end{eqnarray}
\begin{eqnarray}
F_3(|\Delta|)
=\frac{|\Delta|^2}{4H}
&-&\frac{N_c}2\sum_{\pm}\int\frac{d^3p}{(2\pi)^3}\Big [\big |\nu_5+\nu\pm\sqrt{|\Delta|^2+(|\vec p|-\mu_5-\mu)^2}\big|
+\big |\nu_5-\nu\pm\sqrt{|\Delta|^2+(|\vec p|-\mu_5+\mu)^2}\big|\nonumber\\&&
+\big |\nu_5+\nu\pm\sqrt{|\Delta|^2+(|\vec p|+\mu_5+\mu)^2}\big|+
\big |\nu_5-\nu\pm\sqrt{|\Delta|^2+(|\vec p|+\mu_5-\mu)^2}\big|\Big ].
\label{1910}
\end{eqnarray}
(At $\mu_5=0$ these projections have been obtained in Appendix C of Ref. \cite{2color}. The
contribution of the $\mu_5\ne 0$ can be taken into account by the procedure (\ref{71}).
Throughout the paper the integration over the three-momentum $\vec p$ in Eqs.
(\ref{1890})-(\ref{1910}) occurs in the region $|\vec p|<\Lambda= 657$ MeV.)
Then, comparing the smallest values of these functions, we can determine the GMP of the initial TDP
(\ref{72}), and, therefore, the phase in which the system is located in the chiral limit, $m_0=0$, at given
values of chemical potentials. The behavior of the GMP of the model thermodynamic potential (\ref{72}) vs.
chemical potentials supplies us with the full $(\mu,\nu,\nu_5,\mu_5)$-phase portrait of the two-color NJL
model (\ref{1}). Indeed, it is no more than a one-to-one correspondence between any point
$(\mu,\nu,\nu_5,\mu_5)$ of the four-dimensional space of chemical potentials and possible model
phases (CSB, charged PC, BSF and symmetric phase). However, it is clear that this four-dimensional
phase portrait (diagram) is quite bulky and it is rather hard to imagine it as a whole. So in order
to obtain a more deep understanding of the phase diagram as well as to get a greater visibility
of it, it is very convenient to consider different low-dimensional cross-sections of this general
$(\mu,\mu_{5},\nu,\nu_{5})$-phase portrait, defined by the constraints of the form $\nu= const$ or
$\mu_5=const$ and $\nu_5=const$, etc. In the next subsections these different cross-sections of
the most general phase portrait will be presented. But before that, let us discuss the role and
influence of the dual symmetries ${\cal D}_1$ (\ref{55}), ${\cal D}_2$ and ${\cal D}_3$ (\ref{60})
of the model TDP on the shape of its different phase portraits (see also the relevant paper
\cite{2color}).

\subsection{Dual symmetries of the TDP (\ref{72}) and phase portraits of the model at $\mu_5\ne 0$}

As it is clear from the previous subsections, chiral chemical potential $\mu_5\ne 0$ does not
spoil symmetries of the model TDP with respect to duality transformations (\ref{55}) and (\ref{60}),
observed first of all at $\mu_5=0$.

Recall that by duality property (or symmetry, or relation,
etc) of any model, we understand any discrete symmetry of its TDP with respect to
transformations as order parameters (in our case, condensates $M$, $\pi_1$ and $|\Delta|$) and
free external parameters of the system (these may be chemical potentials, coupling constants, etc).
The presence of the dual symmetry of the model TDP means that in its phase portrait there is some
symmetry between phases with respect to the transformation of external parameters, which can
greatly simplify the construction of the full phase diagram of the system. (The invariance of the
TDP (\ref{72}) with respect to sign reversal for any two of the four chemical potentials  is the
simplest example of the dual symmetry of the model (\ref{1}). Due to this kind of  duality, it is
enough to study the phase structure of the model only, e.g., at $\mu\ge 0$,  $\nu\ge 0$,
$\nu_5\ge 0$ and for arbitrary sign of $\mu_5$, etc.) Below, we investigate the phase portrait of
the model (\ref{1}) in the mean-field approximation in the presence of four nonzero chemical
potentials, $\mu$, $\nu$, $\nu_5$, and $\mu_5$ in the chiral limit, putting a special attention to
the role of $\mu_5$ in its formation. In this case, the problem is greatly simplified due to the
fact that the model TDP (\ref{72}) has three dual symmetries, ${\cal D}_1$ (\ref{55}) and
${\cal D}_2$, ${\cal D}_3$ (\ref{60}).

Indeed, let us suppose that $m_0=0$ and that at the point $(\mu=a, \nu=b, \nu_5=c,\mu_5=d)$ of the
phase portrait %of the model
the GMP of the TDP (\ref{72}) lies, e.g., at the point of the condensate space of the form
$(M=A,\pi_1=0,|\Delta|=0)$, i.e. in this case
the CSB phase is realized in the system. Then, according to the symmetries ${\cal D}_2$ and
${\cal D}_3$ (\ref{60}), the TDP has the same meaning if we interchange the values of
chemical potentials and simultaneously appropriately transpose the values of the condensates.
As a result we see that, e.g., at $\mu=c$, $\nu=b$, $\nu_5=a$, $\mu_5=d$ and in the point
$(M=0,\pi_1=0,|\Delta|=A)$ (it is the result of the action of the ${\cal D}_2$ dual transformation
on the TDP (\ref{72})) as well as that at $\mu=a$, $\nu=c$, $\nu_5=b$, $\mu_5=d$ and in the point
$(M=0,\pi_1=A,|\Delta|=0)$ (it is the application of the ${\cal D}_3$ dual transformation
to the TDP) it has the initial meaning. Moreover, it is evident that these new points of the
condensate space are nothing but the GMPs of the TDP (\ref{72}) after its ${\cal D}_2$ and
${\cal D}_3$ transformations (see a more detailed discussion in Ref. \cite{2color}).
Consequently, at the points $(\mu=c, \nu=b, \nu_5=a,\mu_5=d)$ and $(\mu=a, \nu=c, \nu_5=b,\mu_5=d)$
of the phase diagram of the model, which we call dually ${\cal D}_2$ and dually ${\cal D}_3$
conjugated to the starting point $(\mu=a, \nu=b, \nu_5=c,\mu_5=d)$ of the phase portrait, there are
BSF and charged PC phases that are respectively dually ${\cal D}_2$ and dually ${\cal D}_3$
conjugated to the initial CSB phase of the model. Thus, knowing the phase of the model, which is
realized at some point of its phase portrait, we can predict which phases are arranged at the
dually conjugated points of a phase diagram. Moreover, the order parameter of the initial CSB phase
of the point $(\mu=a, \nu=b, \nu_5=c,\mu_5=d)$, i.e. the quantity $M=A$, is equal to the order
parameter $|\Delta|=A$ of the ${\cal D}_2$-dually conjugated BSF phase of the point
$(\mu=c, \nu=b, \nu_5=a,\mu_5=d)$ of the model phase portrait, etc.

At $m_0=0$ each duality transformation ${\cal D}_i$ ($i=1,2,3)$ (\ref{55}) and (\ref{60}) of the
TDP can also be applied to an arbitrary phase portrait of the model as a whole. In particular, it
is clear that if we have a most general $(\mu,\nu,\nu_5,\mu_5)$-phase portrait, then the
action, e.g., of the ${\cal D}_3$ on the TDP can be understood as the following dual ${\cal D}_3$
transformation of the model $(\mu,\nu,\nu_5,\mu_5)$-phase portrait. Namely, it is necessary to
rename both the diagram axes and phases in such a way, that $\nu\leftrightarrow\nu_5$
and CSB$\leftrightarrow$charged PC. At the same time the $\mu$- and $\mu_5$-axes and BSF and
symmetrical phases should not change their names. It is evident that after such ${\cal D}_3$
transformation the $(\mu,\nu,\nu_5,\mu_5)$-phase diagram is mapped to itself, i.e. the most
general $(\mu,\nu,\nu_5,\mu_5)$-phase portrait of the model is self-${\cal D}_3$-dual.
In a similar way it is possible to describe the action of other, ${\cal D}_1$ and ${\cal D}_2$,
duality transformations on the $(\mu,\nu,\nu_5,\mu_5)$-phase portrait of the model, which is,
of course, invariant, or self-dual, under these mappings. But different cross-sections of the
full $(\mu,\nu,\nu_5,\mu_5)$-phase diagram, e.g., the $(\mu,\nu)$-phase portrait at some fixed
values of $\nu_5$ and $\mu_5$, are not invariant, in general, under the action of dual
transformations (see below for some examples). Finally, note that under any ${\cal D}_i$
($i=1,2,3)$ transformation the symmetrical phase remains intact, i.e. it does not change its
position on the phase diagram.

As a result, based on this mechanism of dual transformations of different cross-sections of the full
phase diagram of the two-color NJL model (\ref{1}), it is possible, having a well-known phase
diagram of the model, to obtain its phase portrait in a less studied range of values of chemical
potentials (see some examples from the Ref. \cite{2color}, as well as the following sections of
this paper). In particular, using this approach, one can draw quite definite conclusions about the
effect of chiral asymmetry, i.e. $\mu_5$, on the phase structure of the model (see below).

\section{Phase diagram of the model}

\subsection{Phase structure with $\mu_5\ne 0$ and either one of $\mu$, $\nu$ and $\nu_5$ is nonzero}

Let us start the investigation of the phase structure of the model from a rather simple case when
there is a nonzero chiral imbalance $\mu_{5}$ in quark matter and in addition either one of the
basic chemical potentials, baryon $\mu_B\equiv 2\mu$, isopsin $\mu_I\equiv 2\nu$ or chiral isospin
$\mu_{I5}\equiv 2\nu_5$ chemical potentials, is also nonzero (the value of the remaining two are zero).

First, suppose that only $\mu_{5}$ and $\nu_5$ chemical potentials are nonzero (hence $\mu=\nu=0$).
Then, in the limiting case when $\mu_{5}=0$ and $\nu_5=0$, chiral symmetry of the model (\ref{1}) is
broken spontaneously and there is a dynamical generation of nonzero
quark mass $M$ in the system.

The quark matter with chiral imbalance has been considered in Refs. \cite{ruggieri,andrianov,cao,braguta}.
And it was revealed that if one increases chiral imbalance $\mu_5$, then the chiral symmetry
breaking phenomenon would consolidate, i.e. the CSB phenomenon would be enhanced and dynamical
quark mass $M$ would increase. \footnote{In the chiral limit, the dynamical
quark mass $M$ is the order parameter of the CSB phase. }
This is the so-called effect of catalysis, or enhancement, of chiral symmetry breaking by chiral
imbalance that was observed also in three color QCD, but valid and strictly speaking established
for the first time on lattice just in two color QCD (see in Ref. \cite{braguta}).
Moreover, in Ref. \cite{2color} it was established that in another limiting case, $\mu_{5}=0$ and $\nu_5>0$,
the CSB phase is also realized at any value of $\nu_5$, provided that it is inside the scope of the
validity of the model, $\nu_5<\Lambda$. Starting from the
TDP (\ref{72}) (or comparing the least values of the projections $F_i$ (\ref{1890})-(\ref{1910})),
it is possible to obtain the $(\nu_5,\mu_5)$-phase portrait of the model at $\mu=\nu=0$
(see in Fig. 1 (a)). \footnote{Note that all phase
portraits of Fig. 1 as well as some of the following diagrams are indeed the cross-sections of the
full phase diagram of the model under the constraint that at least one of the four chemical potentials
is zero. In these cases it is enough to consider the phase portrait only at positive values of
chemical potentials. It is a consequence of the fact that if one of the chemical potentials is zero, then
the TDP (\ref{72}) is an even function with respect to each of the remaining nonzero chemical potentials.
So in each phase diagram of Fig. 1 only the regions with $\mu_5>0$ are prepared.}
As it is clear from the figure, in this case just the CSB phase occupies all the physically % accepted as in the scope of applicability
accepted region of chemical potentials, $\mu_5,\nu_5<\Lambda$. Moreover, if one increases any type of chiral imbalances, %any type of chiral imbalances is increased,
i.e. $\mu_5$ or $\nu_5$, then the CSB phenomenon would be enhanced and dynamical quark mass $M$ value would increase.
This could be called the effect of catalysis of chiral symmetry breaking by chiral imbalance of any form, $\mu_5$ or $\nu_5$.

In order to support this statement, in Fig. 2 (a) the behaviour of $M$ as a function of $\mu_5$ at,
e.g., $\nu_5=0.1$ GeV is shown for the two-color NJL model (\ref{1}) at $\mu=\nu=0$. Since this function is an increasing one, we can conclude that CSB is enhanced (or catalysed) by $\mu_5$ in this case. The behaviour
of $M$ vs $\mu_5$ at other fixed values of $\nu_5$ is similar. Moreover, in this case, i.e. at
$\mu=\nu=0$, the order parameter $M$ vs $\nu_5$ at different fixed $\mu_5$ values is also an
increasing function, and in particular at $\mu_5=0.1$ GeV the plot of this function coincides with
a curve drawn in Fig. 2 (a) in which one should rename $\mu_5$ axis by $\nu_5$. (It follows from the fact that at $\mu=\nu=0$ the projection $F_1(M)$ (\ref{1890}) is symmetrical under the replacement
$\mu_5\leftrightarrow\nu_5$.) So increasing at
$\mu=\nu=0$ either one of chiral imbalances, i.e. chemical potentials $\mu_5$ or $\nu_5$, or both
of them simultaneously, the CSB phase only solidifies. In this case the $(\nu_5,\mu_5)$-phase
structure is not complicated, it is CSB phase everywhere for all physically accepted values of
$\mu_5,\nu_5$.

This effect could be explained with the following qualitative arguments. The Fermi energies 
(\ref{1000}) of
$u_{L}$ and $u_{R}$ quarks in this case have a rather simple form $\mu_{u_{L}}=\nu_5+\mu_5$ and
$\mu_{u_{R}}=-\nu_5-\mu_5$. If $\nu_5=0$ then $\mu_{u_{L}}=\mu_5$, $\mu_{u_{R}}=-\mu_5$ and
condensation of $\overline{u}_{R}u_{L}$ is quite feasible and this leads to chiral symmetry 
breaking. One could note also that if $\mu_5$ is increased, the number of states at Fermi 
spheres are getting larger and the value of the condensate is increased. The same effect could 
be observed in the $\mu_5=0$ case and the effect is fully identical, $\mu_{u_{L}}=\nu_5$ and 
$\mu_{u_{R}}=-\nu_5$. The effect of CSB catalysis simultaneously by both $\mu_5$ and $\nu_5$ is nicely described by 
these qualitative reasoning as well.

Now if there is an isospin imbalance, $\nu\ne 0$, together with chiral imbalance,
$\mu_5\ne 0$, in the system (hence, we suppose that other chemical potentials are zero,
$\mu=\nu_5=0$), then the situation is drastically different. In this case, to obtain the
$(\nu,\mu_5)$-phase portrait of the massless two-color NJL model (\ref{1}) (see in Fig. 1(b)), it
is enough to apply, without any numerical calculations, to the phase diagram of Fig. 1(a) the dual
${\cal D}_3$ transformation. The corresponding mechanism is described in Sec. III C, so we should
change in Fig. 1 (a) $\nu_5\to\nu$ and CSB $\to$ charged PC.
As a result, we have a phase portrait of Fig. 1 (b). (Recall that the particular case of this
diagram, when $\nu>0$ and $\mu=\nu_5=\mu_5=0$, was discussed in Ref. \cite{2color}, where it was
established that at the points of the $\nu$ axis of the diagram the charged PC phase is arranged
at least for $\nu<\Lambda$. It was known that the order parameter of this phase,
i.e. the charged pion condensate $\pi_1$, is enlarged if isospin density, i.e. isospin chemical
potential, grows. It turns out that the pion condensate is also an increasing function vs $\mu_5$.
It is an extremely surprising feature that $\mu_5$ completely changed gears and, once isospin
density is nonzero, start to catalyze charged pion condensation phenomenon instead of chiral symmetry
breaking one.  It is a consequence of the fact that in this case the condensate $\pi_1$ is the
${\cal D}_3$ mapping of the chiral condensate $M$ of the CSB phase in Fig. 1 (a). Hence, for the
charged PC phase of Fig. 1 (b) the plot of its order parameter $\pi_1$ vs $\mu_5$ at, e.g.,
fixed $\nu=0.1$ GeV can be obtained
from Fig. 2 (a) by two simple replacements, $M\to\pi_1$ and $\nu_5=0.1$ GeV $\to$ $\nu=0.1$
GeV, etc. So in this particular case when $\mu_5\ne 0$, $\nu\ne 0$ but $\mu=\nu_5= 0$, the chiral
imbalance $\mu_5$ serves as a factor that catalyses (or enhances) the spontaneous breaking of the
isospin $U(1)_{I_3}$ symmetry, which manifests itself (without $\mu_5$) even at $\nu > 0$,
$\mu=\nu_5=\mu_5= 0$.

Now let us make a couple of comments on the chiral limit and the current quark masses. Figs. 1
have been shown in the chiral limit, i.e. at zero current quark mass $m_0$. At the physical point,
at physical value of the current quark mass, the $(\nu_5,\mu_5)$-phase diagram of Fig. 1(a) would be 
the same, whereas the $(\nu,\mu_5)$-phase portrait of Fig. 1 (b) would slightly change. One can see that
pion condensation in Fig. 1 (b) starts at infinitesimally small values of isospin chemical
potential $\nu$ in the chiral limit. In this case pion mass is zero and isospin density becomes nonzero 
at infinitesimally small values of isospin. In the physical point, at physical pion mass, isospin 
density emerges at isospin chemical
potential $\nu$ reaching the value of half of the pion mass, so one would see a small region of CSB phase
at $\nu<m_{\pi}/2$ and only at $\nu>m_{\pi}/2$ charged PC phase sets in. So the $\mu_5$
catalysis of chiral symmetry breaking will be switched to the catalysis of charged PC
phenomenon at  $\nu=m_{\pi}/2$, so overall the picture would stay the same.
\begin{figure}
%----figure 1
\hspace{-1cm}\includegraphics[width=1.0\textwidth]{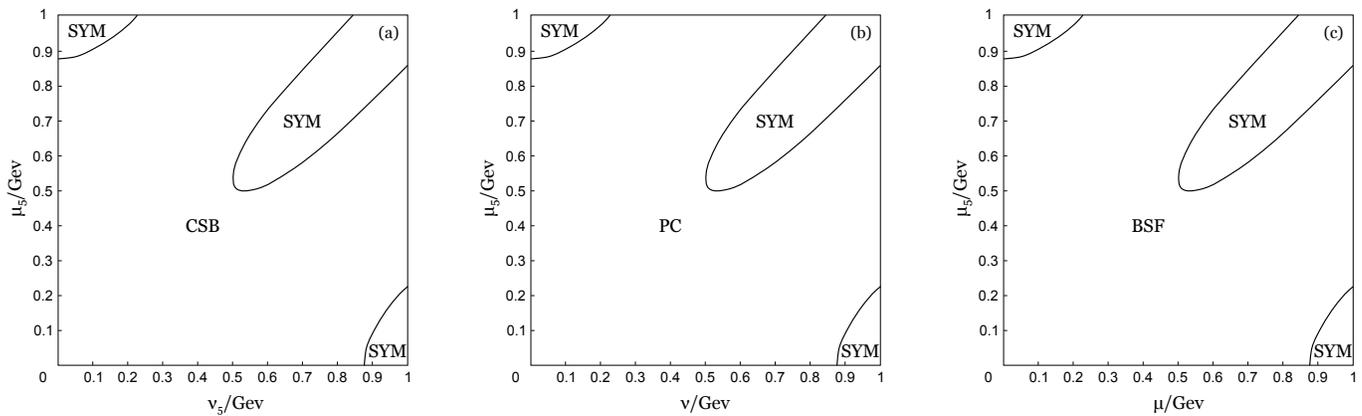}
 %\hfill
%\includegraphics[width=0.35\textwidth]{fig2.pdf}
%\hfill
%%\includegraphics[width=0.35\textwidth]{fig3.pdf}\\
%\parbox[t]{0.45\textwidth}{
 \caption{ (a) $(\nu_5,\mu_5)$-phase
 diagram at $\nu=\mu=0$ GeV.  (b) $(\nu,\mu_5)$-phase diagram
 at $\mu=\nu_5=0$ GeV. (c) $(\mu,\mu_5)$-phase diagram at $\nu_{5}=\nu=0$ GeV. In these diagrams BSF
 means the baryon superfluid phase, PC -- the charged pion condensation phase, CSB -- the chiral
 symmetry breaking, and SYM denotes the symmetrical phase of the model.
}
\end{figure}

Finally, if we are going to consider the phase structure of the two-color NJL model (\ref{1}) in
the chiral limit and in the case when $\mu$- and $\mu_5$-chemical potentials are nonzero but
$\nu=\nu_5=0$, it is sufficient to perform the ${\cal D}_2$-dual transformation of the phase
portrait of Fig. 1 (a), or, alternatively, the ${\cal D}_1$-dual transformation of the phase
portrait of Fig. 1 (b). The results of these dual mappings is the $(\mu,\mu_5)$-phase diagram of
Fig. 1 (c), in which for the whole range of values of chemical potentials the baryon superfluid phase is
arranged. The order parameter $|\Delta|$ of this phase is also an increasing function vs $\mu_5$
and/or $\mu$ (it is clear after dual ${\cal D}_2$ transformation of Fig. 2 (a), i.e. replacing
there $M$ $\to$ $|\Delta|$ and the constraint $\nu_5=0.1$ GeV $\to$ $\mu=0.1$ GeV). Hence, in this
region of chemical potentials the $\mu_5$ catalyses/increases the BSF phase and spontaneous
breaking of the baryon $U(1)_B$ symmetry.  The fact that $\mu\ne 0$ leads to diquark
condensation is natural and well known, whereas the same effect on phenomenon of diquark pairing
by chiral imbalance is rather curious. This catalysis effect could be comprehended just by
arguments of pairing on the Fermi surface, with the following qualitative arguments. 

One can see from Eq. (\ref{1000}) that the
Fermi momenta of $u_{L}$ and $d_{L}$ quarks in this case, i.e. at $\nu=\nu_5=0$, are 
$\mu_{u_{L}}=\mu+\mu_5$ and
$\mu_{d_{L}}=\mu+\mu_5$. As a result, it is clear that the condensation of Cooper $u_{L}d_{L}$ 
pairs and appearing of the BSF phase
is possible, in particular, both at $\mu\ne 0$, $\mu_5=0$ and at  $\mu=0$, $\mu_5\ne 0$. 
But in the last case we have from Eq. (\ref{1000}) that $\mu_{u_{L}}=\mu_5$ and
$\mu_{u_{R}}=-\mu_5$, i.e. an equal possibility for the condensation of the 
$\overline{u}_{R}u_{L}$ pairs and generation of the CSB phenomenon. However, if both 
$\mu\ne 0$ and $\mu_5\ne 0$, then Fermi energies of $u_{L}$ and 
$d_{L}$ quarks are equal (see above), which favors the formation of $u_{L}d_{L}$ Cooper pairs,
but $\mu_{u_{L}}=\mu+\mu_5$ and $\mu_{u_{R}}=\mu-\mu_5$. 
So there is a mismatch in the Fermi surfaces of $\overline{u}_R$ and $u_L$ quarks which obstructs the 
formation of the Cooper $\overline{u}_{R}u_{L}$ pairs. This allows us to say that, at 
$\mu\ne 0$ and $\mu_5\ne 0$, the formation of the BSF phase is preferable to the generation of the CSB phenomenon. 
Finally, we see that Fermi energies of $u_{L}$ and $d_{L}$ quarks (recall, in this case they are $\mu+\mu_5$) are a 
growing vs $\mu_5$ quantities, hence the order parameter $|\Delta|$ of the BSF phase is also increases with $\mu_5$, i.e. there is a catalysis of 
the BSF by chiral imbalance $\mu_5$. 

So from the whole picture of the phase structure of chirally imbalanced, $\mu_5\ne 0$, two-color
quark matter that has another additional chemical potential, one can infer the following inherent
characteristics of $\mu_5$. Chiral imbalance $\mu_5$ enhances/catalyses any symmetry breaking
phenomena realized in the system when it is under the influence of only this additional chemical
potential. For example, if only $\mu\ne 0$, then baryon $U(1)_B$ symmetry is broken spontaneously
(diquarks have a nonzero condensate $|\Delta|$) and $\mu_5$ enhances just this effect. If there is
a nonzero isospin imbalance in the system (only $\nu\ne 0$) and charged PC phenomenon is observed
in such a quark matter at $\mu_5=0$, then nonzero $\mu_5$ (together with nonzero $\nu$) enhances
just the charged PC phenomenon as well as the spontaneous isospin $U(1)_{I_3}$ symmetry breaking.
Finally, if  two-color quark matter is characterized by chiral isospin imbalance, only
$\nu_5\ne 0$, then at $\mu_5=0$ the axial isospin (or chiral) $U(1)_{AI_3}$ symmetry is broken
spontaneously, CSB phase is realized in the system and quarks acquire dynamically a nonzero mass
$M$. And in this case the nonzero $\mu_5$ catalyses/enhances just this effect. So one can say that
chiral imbalance $\mu_5$ is a universal catalyzer for every phenomena in two color quark matter.
But it cannot trigger any phenomenon itself. It is in a way as chameleon, which assumes the color
of the environment, it only enhances the phenomenon picked by other conditions of the medium.

Although quite surprising and curious, this whole picture is the direct consequence of the duality
properties of the phase diagram. At $\mu_5=0$ the ($\mu$,$\nu$,$\nu_5$)-phase diagram of the
two-color NJL model (\ref{1}) is highly symmetric due to the dualities (see in Ref. \cite{2color}),
and the expansion of this phase diagram to nonzero values of chiral chemical potential is also 
highly symmetric %also possesses this
%high symmetry wts 
(see in the section III B) and possesses this high dual symmetry as well. Now, for example,
if one imagines the full ($\mu$,$\nu$,$\nu_5$,$\mu_5$)-phase diagram as a cube elongated to the
$\mu_5$ direction, then one can pick three facets of this phase diagram, namely ($\nu_5=0$ and
$\nu=0$), ($\nu=0$ and $\mu=0$) and ($\mu=0$ and $\nu_5=0$), these facets are ($\mu$,$\mu_5$),
($\nu_5$,$\mu_5$) and ($\nu$,$\mu_5$) cross-sections considered above in Fig. 1. So due to
preservation of
dualities at $\mu_5\neq0$, these facets are dual to each other. If one knows the behavior of the
phase structure in one of them, then the others are predetermined. In principle, one could get a
lot of hints that CSB is catalyzed by chiral imbalance $\mu_5\neq0$ and $\nu_5\neq0$ in the whole
plane ($\nu_5$,$\mu_5$) of Fig. 1(a). First, one knows that it was shown separately for each chiral chemical
potential, and also it was bolstered by qualitative pairing arguments. Second, these arguments are
easily generalized to the case of both simultaneously nonzero $\mu_5$ and $\nu_5$.
Then dualities dictate that chiral imbalance $\mu_5$ leads to the catalysis of charged pion
condensation at $\nu\neq0$, see Fig. 1 (b), and that it catalyzes diquark condensation at $\mu\neq0$, see Fig. 1 (c).
So if it is obtained that chiral imbalance $\mu_5$ catalyzes some phenomenon in some cross-section
of the full phase diagram, then the other phenomena are catalyzed by $\mu_5$ in dually conjugated
cross-sections. The feature of chiral imbalance $\mu_5$ of having the chameleon property is %an
%implication of phase diagram having duality properties,
a consequence of dual properties of the
full phase diagram of the model.

Let us reiterate the arguments,  the whole ($\mu,\nu,\nu_5,\mu_5$)-phase diagram is
interconnected and highly symmetric due to dualities. Chiral chemical potential $\mu_5$ is not involved in duality transformations and in this sense it stands alone
from other chemical potentials. The $\mu_5$ is not involved in it that much,
but the influences of $\mu_5$ on various phenomena are interconnected, and in this way the
influences of $\mu_5$ on the ($\mu,\nu,\nu_5$)-phase diagram is partly constrained
by duality properties. Just from the known facts that chiral imbalance $\mu_5$ as well as $\nu_5$ is a catalyst of chiral symmetry
breaking and the existence of dual properties of phase diagram, one could have inferred that
chiral imbalance $\mu_5$ has a property of being {\it universal catalyst} in two-color quark matter.
\begin{figure}
%----figure 2
\hspace{-1cm}\includegraphics[width=1.0\textwidth]{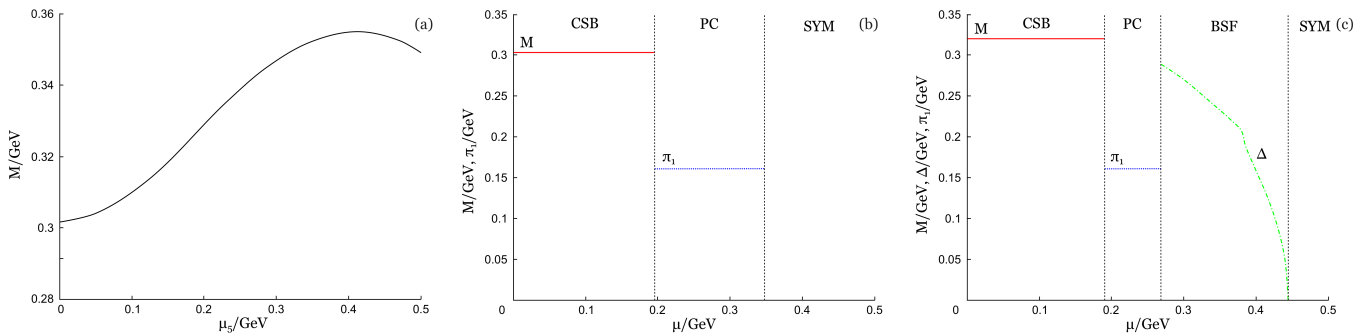}
%\hfill
%\includegraphics[width=0.34\textwidth]{M_pi_D_vs_mu_nu5_024_mu5_05.eps}
%\hfill
%\includegraphics[width=0.34\textwidth]{M_pi_D_vs_mu_nu5_02_mu5_05.eps}\\
%\parbox[t]{0.45\textwidth}{
 \caption{ The behaviors of condensates vs. chemical potentials: (a) $M$ as a function of  $\mu_5$
 at $\nu_5=0.1$ GeV, $\mu=\nu=0$ GeV.
 (b) $M$ and $\pi_1$ as a functions of $\mu$ at $\nu=0$, $\nu_{5}=0.25$ GeV and $\mu_5=0.5$ GeV
 (c) $\Delta$, $\pi_1$ and $M$ as a functions of $\mu$ at $\nu=0$, $\nu_{5}=0.2$ GeV and $\mu_5=0.5$ GeV.
 The notations are the same as in Fig. 1.
}
\end{figure}

Let us make one more comment. One can note that in this case of only one nonzero basic chemical
potential, there exists the effect that each phenomenon is tightly connected %by one-to-one correspondence 
with the chemical potential that it is triggered by. This feature, noticed in the
case of $\mu_5=0$, is transferred to the case $\mu_5\neq0$.

\subsection{Chemical potential $\mu_5$ as a catalyst of different phenomena in more general cases}

In this section let us consider the situation when basic chemical potentials, $\mu$, $\nu$ and
$\nu_5$, reaches rather moderate values.

%(quantitatively it is to be discussed in a second).
Let us commence with the
situation when only two from basic triple of chemical potentials are nonzero and at least one of them
is smaller than 0.2 GeV. For instance, first consider the quark matter with nonzero baryon density,
$\mu\neq 0$, and isospin imbalance, $\nu\neq0$, and without chiral imbalances, i.e. $\mu_5=\nu_5=0$.
Recall that in the particular case when $\mu=\mu_5=\nu_5=0$ and $\nu\neq0$
the charged pion condensation is realized (see in Fig. 1 (b)), and this passes to the slice of charged PC phase at
rather large $\nu$ at ($\mu$, $\nu$)-phase diagram (see Fig. 3 (a)).
But if the isospin imbalance is not so large ($\nu$ smaller than 0.2 GeV), then at baryon density corresponding to the value $\mu=\nu$ the phase transition to BSF phase takes place and diquark
condensation appears in the system, and at rather large $\mu>\nu$ there is a whole slice of BSF
phase (see Fig. 3 (a)). So here, i.e. in the case of only two  nonzero basic chemical potentials,
if one talks about the fact that basic chemical potentials have rather moderate values,
it means that at least one of them is smaller than 0.2 GeV and the other one could be quite large.
One could behold that in this regime the largest chemical potential defines the phase
structure: if it is baryon chemical potential then diquark condensation prevail, and if it is
isospin one then charged pion condensation takes over.

Now, if one increases the value of chiral chemical potential $\mu_5$, which was equal to zero in
the discussion above, then the ($\nu$, $\mu$)-phase diagram of Fig. 4 (a) at $\mu_5=0.4$ GeV and
$\nu_5=0$ could be obtained by numerical analysis, and one can note that in the discussed regime
(moderate values of baryon and isospin chemical potentials or at least one of them
is smaller than 0.2 GeV) the phases and the phase transition do not alter at all (see in Fig. 4 (a)
the region of small values, $<$ 0.2 GeV, of $\mu$ or $\nu$).

Now let us turn our sight to the other cross-sections of the full phase diagram containing chiral
chemical potential $\mu_5$. In particular, it is possible to consider how the
$(\nu,\mu_5)$-phase diagram of Fig. 1 (b), which is drawn under the condition $\mu=\nu_5=0$, changes when
one increases $\mu$ up to 0.2 GeV ($\nu_5$ is still equal to zero). Or, in other words, it is an expansion
of the cross-section of fixed $\mu$ to nonzero values of $\mu_5$. Numerical analysis shows that in
this case the typical $(\nu,\mu_5)$-phase portrait of the model looks like the one of Fig. 5 (a),
which is depicted for fixed $\mu= 0.1$ GeV. It means that in these phase portraits at $\nu<\mu$ and
$\mu_5>0$ the BSF phase is arranged, but at $\nu>\mu$ and $\mu_5>0$ one can see the charged PC
phase. Interestingly enough that the transition line
between the PC and BSF phases is a straight line (it does not depend on the value of $\mu_5$), and
the phase in this case is entirely chosen by the relation between values of $\mu$ and $\nu$ chemical
potentials. It is in full agreement with the fact that for moderate $\mu$ and $\nu$ values
the phase transitions at the $(\nu,\mu)$-phase diagram of Fig. 4 (a) was not influenced a lot by chiral chemical potential
$\mu_5$. Bearing in mind that the section of $\mu_5=0$ in Fig. 5 (a) is the section $\mu=0.1$ GeV
in Fig 4 (a), one could easily comprehend that the phase transition in Fig. 5 (a) is at $\nu=\mu$
and it shifts to larger values of $\nu$ if quark number chemical potential $\mu$ is increased.
Apart from the fact that the transition line in Fig. 5 (a) between the PC and BSF phases is a
straight line, let us elaborate more on the effect of chiral $\mu_5$. As for $\mu_5=0$ as well as
for $\mu_5\neq0$ the same phase structure is observed, i.e. at $\nu<\mu$
($\nu>\mu$) the BSF (the charged PC) phase is realized, but the order parameters of these phases
are growing functions vs $\mu_5$, and the increase is quite dramatic as one can see in Fig. 2 (a).
From all these one can conclude that for
this set of chemical potentials the $\mu_5$ catalyzes (or enhances) the phenomena observed at
$\mu_5=0$. This feature was already noticed and discussed above in Sec. IV A for the particular case of two
nonzero chemical potentials (one of them is $\mu_5$).

\begin{figure}
%----figure 3
\hspace{-1cm}
\includegraphics[width=1.0\textwidth]{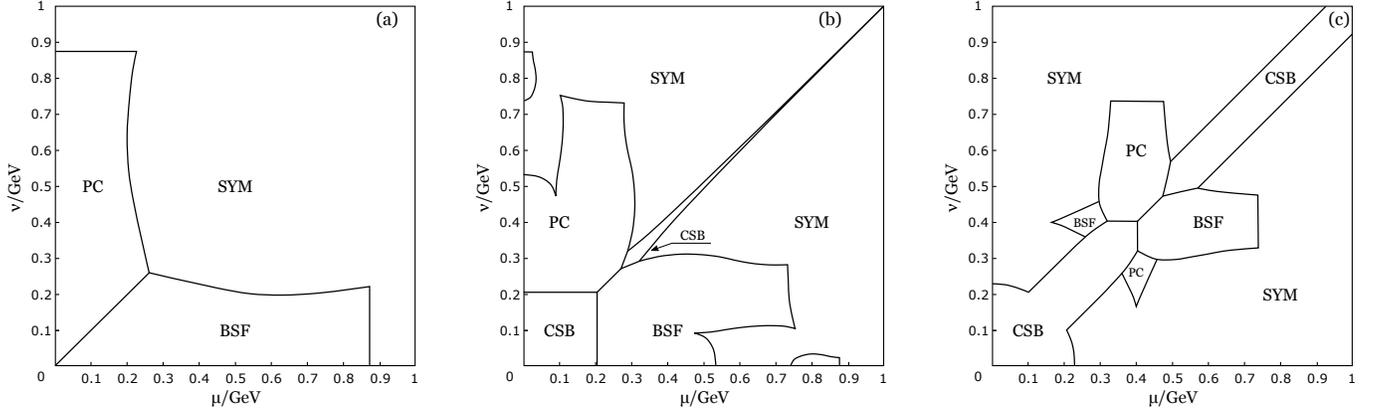}
%\includegraphics[width=0.33\textwidth]%{figmunu02.eps%munu_mu5_04.eps%
%}
%\includegraphics[width=0.33\textwidth]{fignu5.eps}
 %\hfill
%\includegraphics[width=0.35\textwidth]{fig2.pdf}
%\hfill
%%\includegraphics[width=0.35\textwidth]{fig3.pdf}\\
%\parbox[t]{0.45\textwidth}{
 \caption{$(\mu, \nu)$-phase diagrams at: (a) $\mu_{5}=\nu_5=0$ GeV. (b) at $\mu_{5}=0$ GeV,
 $\nu_5=0.2$ GeV. (c)  at $\mu_{5}=0$ GeV, $\nu_5=0.4$ GeV. The notations are the same as in Fig 1.
}
\end{figure}

Applying, for example, to the phase diagram of Fig. 5 (a) the dual ${\cal D}_1$ or dual
${\cal D}_3$ mappings, etc,
one can obtain the $(\mu,\mu_5)$-phase portrait at fixed $\nu_5=0$ and $\nu= 0.1$ GeV (see in
Fig. 5 (b)), or the $(\nu_5,\mu_5)$-phase portrait (see in Fig. 5 (c)) at $\mu=0.1$ GeV, $\nu=0$,
respectively. Since under the dual transformations the properties of the order parameters are not
changed, we can conclude that for these particular sets of the basic chemical potentials the order
parameter of each non-symmetrical phase is indeed an increasing function vs $\mu_5$, i.e. the
chemical potential $\mu_5$ is a catalyst of the phase structure observed at $\mu_5=0$. Generalizing
these observations, we can state that if one of the basic chemical potentials is zero, but the
others are nonzero and at least one of them is not so large, say $<0.2$ GeV, then $\mu_5$
{\cal catalyses} the phenomena that observed in quark matter at $\mu_5=0$.

Now let us turn our gaze to the phase structure of the model in the case of three nonzero basic
chemical potentials (and first keep the value of chiral imbalance $\mu_5$ equal to zero) in the
regime of moderate values, when at least two of them are smaller than 0.2 GeV. In this case the
($\mu$, $\nu$)-phase diagram at $\nu_5=0.2$ GeV is depicted in Fig. 3 (b) and one can see that if
$\mu=0$ then at $\nu<\nu_5=0.2$ GeV the system is in CSB phase as it should be since there is
rather large chiral imbalance $\nu_5$. Then, if value of $\nu$ is increased and reaches the value
of $\nu=\nu_5=0.2$ GeV, then the phase transition to the charged PC phase takes place and at
$\nu>\nu_5=0.2$ GeV the charged pion condensation only solidifies. Now let us take $\nu=0$.
As it clear from Fig. 3 (b), at $\mu<\nu_5=0.2$ GeV there is CSB phase due to the same reasons. But at $\mu=\nu_5=0.2$ GeV the phase transition
to BSF phase occurs, and diquark condensation is prevailing. In the latter example, if one assume
that $\nu\neq0$ but less than $\nu_5=0.2$ GeV then nothing changes and the phase transition occurs at exactly the same point $\mu=\nu_5=0.2$ GeV, so one can conclude that if $\nu<\mu, \nu_5$ then it does not influence much the phase structure at all. Now, if $\nu>\nu_5=0.2$ GeV then one can see in Fig. 3 (b) that at increasing $\mu$ the charged PC phase continues to dominate up to the value of $\mu=\nu$, and $\nu_5$ does not play almost any role here.

So, at $\mu_5 =0$ one can characterize the general picture (in the case of moderate
basic chemical potential values) in the following manner. (i) Each basic chemical potential (i.e.
$\mu$, $\nu$ or $\nu_5$) is in
one-to-one correspondence to one of CSB, BSF or charged PC phases, i.e. to a corresponding
condensation and symmetry breaking patterns. For example, $\mu$ corresponds to BSF phase,
$\nu$ -- to charged PC, and $\nu_5$ -- to CSB phase. (ii) The largest chemical potential sets
the corresponding phase that occupies the
system. Hence, if the triplet of basic chemical potentials has moderate values and $\mu_5 =0$, then to get the
phase structure one could find just the basic chemical potential with the highest value, and it
defines the phase that is realized in the system. The other two do not play a significant role here.
This picture unfolds only in the regime of moderate values of basic chemical potentials (recall,
by this regime we mean here that at least two from three basic chemical potentials have values
smaller than 0.2 GeV).  Let us stress that we have chosen the value $\nu_5=0.2$ GeV in Fig. 3 (b)
on purpose. Based on it one can easily envisage, for example, the phase diagram at $\nu_5=0.1$ GeV,
since it is a borderline value and one could see that if $\mu_5$ is also small (in Fig. 3 (b)
$\mu_5=0$), then this picture holds even if values of $\mu$ and $\nu$ can reach almost $0.3$ GeV,
even slightly out of the scope that we called moderate values. But if the values of  $\mu$, $\nu$
and $\nu_5$ lies inside the moderate regime, i.e. at least two of them are less than $0.2$ GeV,
then this concise and elegant picture inhesitantly persists to any values of chiral chemical
potential $\mu_5$, one could see this in Figs. 5.

\begin{figure}
%----figure 4
\hspace{-1cm}
\includegraphics[width=1.0\textwidth]{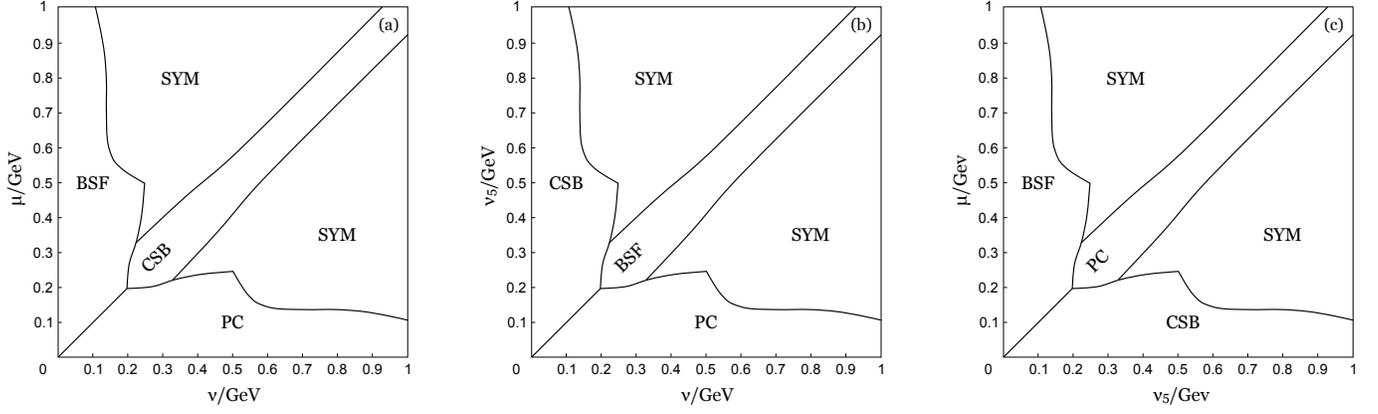}
%\includegraphics[width=0.33\textwidth]{munu_mu5_04.eps%fignunu5_04.eps
%}
%\includegraphics[width=0.33\textwidth]{fignunu5_04.eps%munu_5_mu5_04.eps
%}
%\includegraphics[width=0.33\textwidth]{munu_5_mu5_04.eps%fignunu5_04.eps%munu_mu5_04.eps
%}
 %\hfill
%\includegraphics[width=0.35\textwidth]{fig2.pdf}
%\hfill
%%\includegraphics[width=0.35\textwidth]{fig3.pdf}\\
%\parbox[t]{0.45\textwidth}{
 \caption{ Dually conjugated phase portraits: (a) $(\nu, \mu)$-phase diagram at $\mu_{5}=0.4$ GeV, $\nu_5=0$ GeV. (b) $(\nu, \nu_5)$-phase diagram at $\mu_{5}=0.4$ GeV, $\mu=0$ GeV.
 It is a ${\cal D}_2$ mapping of Fig. 4 (a). (c) $(\nu_5,\mu)$-phase diagram at $\mu_{5}=0.4$ GeV, $\nu=0$ GeV. It is a ${\cal D}_3$ mapping of the
 diagram Fig. 4 (a). The notations are from Fig. 1.
}
\end{figure}

Now let us allow chiral chemical potential $\mu_5$ to be nonzero in the case of three nonzero
basic chemical potentials $\mu$, $\nu$ and $\nu_5$. This could be contemplated as the
expansion of the phase diagram portrayed in Fig. 3 (b), where $\mu_5$ was equal to zero, to nonzero
values of $\mu_5$. In the most concise way it could be formulated that in the regime of moderate
values discussed above the phases and phase transitions stay at their places and just condensates
are enhanced, so chiral imbalance $\mu_5$ catalyses all three phases.  It could be demonstrated
probably in the most convenient way by making the value of the fourth chemical potential nonzero,
namely $\nu_5$ in Fig. 5 (a) and increasing its value. If $\nu_5$ is smaller than $\mu=0.1$ GeV
then at $\nu<0.1$ GeV the BSF phase is intact and continues to occupy the whole region of $\nu<0.1$
GeV and the whole phase diagram in Fig. 5 (a) is unchanged and chiral chemical potential $\mu_5$
catalyzes BSF and PC phases. But if the value of $\nu_5>\mu=0.1$ GeV, then the whole region of
$\nu<0.1$ GeV is occupied by CSB phase and chiral chemical potential $\mu_5$ starts to catalyze
chiral condensation in the region $\nu<0.1$ GeV. And if $\nu>0.1$ GeV it continues to catalyze
charged
pion condensation phenomenon. A rather peculiar behavior is observed when $\nu_5=\mu=0.1$ GeV.
Then the whole region $\nu<0.1$ GeV contains two degenerate phases, BSF and CSB, and they are
equally favoured, it is a rather regular thing in, for example, first order phase transitions,
but here this regime occupies the whole phase region not just the line of phase transition and
it is rather peculiar. It could be explained by the following argument. It is actually the phase
transition between BSF and CSB phases that takes place at $\nu_5=\mu>\nu$. But since the value of
the imbalance with the smallest value $\nu$ does not influence the phase structure, this kind of phase transition expands to the
whole region of $\nu<0.1$ GeV values. Let us note that the axis of $\mu_5=0$ in Fig. 5 (a)
corresponds to the line $\mu=0.1$ GeV in phase diagram obtained from the one in Fig. 3 (b) but
drawn at $\nu_5=0.1$ GeV, and one can notice that the line up to $\nu<0.1$ GeV is the phase
transition between BSF and CSB phases. By the way, in this case both degenerate phases, as BSF as
well as CSB, are catalyzed by chiral imbalance $\mu_5$ by the same degree and the picture expands
also to the plane of chiral chemical potential $\mu_5$. Let us also note here that the phase
diagram presented in Fig. 5 (a) (recall that this figure is shown in the chiral limit $m_{0}=0$) does not change in the physical point, when one takes physical
nonzero value of current quark mass, since the value of the quark number chemical potential
$\mu$ is already
larger than half of pion mass.

Concluding, let us reiterate the phase picture in the regime of moderate and rather low values of
basic chemical potentials in a concise form.
It can be characterized in the following manner. Each basic chemical potential is in one-to-one
correspondence with one of the possible nontrivial phases of the model (charged PC, BSF or CSB),
as well as with its order parameter/condensate (isospin chemical potential $\nu$ corresponds to
the charged PC phase, quark number chemical potential $\mu$ -- to the BSF phase, etc) and the
largest chemical potential fully determines the phase that occupies the system. Chiral chemical
potential $\mu_5$ does not influence the phase transition features, in other words $\mu_5$ does
not choose any phase but it {\it catalyzes} the phases, enhancing their condensates, picked by
basic chemical potentials.

\begin{figure}
%----figure 5
\hspace{-1cm}\includegraphics[width=1.0\textwidth]{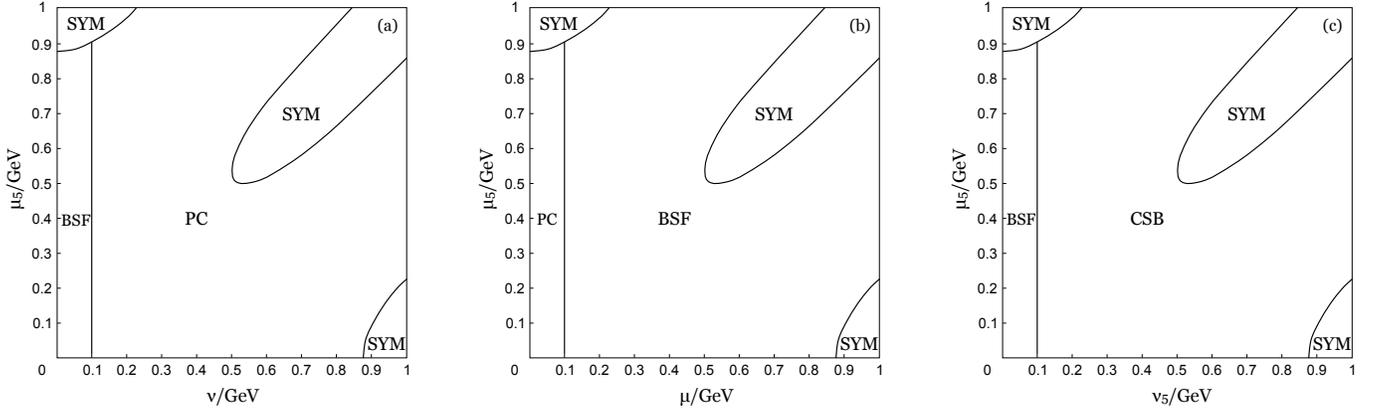}
% \hfill
%\includegraphics[width=0.33\textwidth]{figmu_5_nu_0.1.eps}
%\hfill
%\includegraphics[width=0.33\textwidth]{figmunu02.eps%figmunu02.eps%fignunu502.eps
%}\\
%\parbox[t]{0.45\textwidth}{
 \caption{ Dually conjugated phase portraits: (a) $(\nu,\mu_5)$-phase diagram at $\mu=0.1$ GeV, $\nu_5=0$ GeV. (b) $(\mu,\mu_5)$-phase diagram at $\nu_{5}=0$, $\nu=0.1$ GeV. It is a ${\cal D}_1$ mapping of Fig. 5 (a). (c) $(\nu_5 ,\mu_5)$-phase
 diagram at $\mu=0.1$ GeV, $\nu=0$ GeV. It is a ${\cal D}_3$ mapping of Fig. 5 (a). The notations are from Fig. 1.
}
\end{figure}

\subsection{The ability of $\mu_5$ to generate CSB, charged PC and BSF phenomena}

\subsubsection{The case when two of $\mu$, $\nu$ and $\nu_5$ have rather large values. }

Now let us switch gears a little bit and turn our attention to the regime when basic triplet of
chemical potentials has rather large values. It is reasonable to start with the situation if two of basic chemical potentials are nonzero and rather large but the remaining chemical potential is rather small or even zero.

So take a look at Fig. 3 (a), where  $(\mu, \nu)$-phase diagram at $\mu_5=0$ and $\nu_5=0$ is
depicted. It does not contain any region with CSB phase. Then, with increasing $\mu_5$, when
it reaches the critical value $\mu_{5c}\in 0.15\div 0.2$ GeV, on the $(\mu, \nu)$-phase diagram, which is under the constraint $\nu_5=0$, the band of CSB phase {\it suddenly} appears along the straight line $\mu=\nu$,
when these chemical potentials are larger than $0.2$ GeV. The typical $(\mu, \nu)$-phase diagram at
$\nu_5=0$ and $\mu_5=0.4$ GeV is given in Fig. 4 (a), which is obtained by numeric calculation. And one can see in this figure that for the
values of $\mu\approx\nu$ %, when these chemical potentials are larger than $0.2$ GeV,
the band of the CSB phase is realized (curiously enough that this start to happen at the $\mu=\nu$ equals around $0.2$ GeV).

One can also recall that the same
pattern was observed in the case of nonzero $\nu_5$ at $\mu_5=0$. For example, the
$(\mu, \nu)$-phase diagram at $\nu_5=0.4$ GeV and $\mu_5=0$ is presented in Fig. 3 (c). (It is also
interesting to note that, like any $(\mu, \nu)$-phase diagrams of the model at fixed values of $\nu_5$
and $\mu_5$, the phase portraits in Figs. 3 are self-dual with respect to the
action of the duality-${\cal D}_1$ transformation (\ref{55}) on them, i.e. charged PC and BSF
phases are located on the figures mirror-symmetrically relative to the straight line $\mu=\nu$.)
Comparing the diagrams of Fig. 4 (a) and Fig. 3 (c), one could notice that at rather large values
of $\mu$ and $\nu$ and if $\nu_5$ and $\mu_5$ is
greater than 0.2 GeV, the band of CSB phase,
also the corresponding chiral condensate $M$ and its behaviour vs. chemical potentials does not depend on which chiral chemical potential is nonzero, $\nu_5$ or $\mu_5$. These CSB-bands coincide in the whole range of values of $\mu_5$ and $\nu_5$
and generation of this band-like shaped domain of CSB phase is identical whether the chiral or
chiral isospin imbalance is nonzero, i.e. chiral $\mu_5$ or chiral isospin $\nu_5$ imbalances stand on exactly the same footing in regards of affecting the CSB phase. And also it can serve as an additional indication of the tendency that chiral imbalance
promotes chiral symmetry breaking. But in this case the promotion is qualitatively different from
the one discussed above, and the chiral imbalances {\it genuinely generates} chiral symmetry
breaking and not just enhance (or catalyze) the CSB triggered by other factors.

This behavior, i.e. the generation of band-shaped domain of CSB phase, can be easily qualitatively
understood. As it follows from Eq. (\ref{1000}), for nonzero values of $\nu_5$ in the region 
$\mu\approx\nu$ and $\mu_5=0$ the Fermi momenta of
$d_{L}$ and $d_{R}$ quarks are $\mu_{d_{L}}=\mu-\nu-\nu_5\approx -\nu_5$ and
$\mu_{d_{R}}=\mu-\nu+\nu_5\approx +\nu_5$, respectively. So the condensation of the $\overline{d}_{L}d_{R}$ 
pairs is possible and chiral symmetry is broken down. For nonzero $\mu_5$ and $\nu_5=0$ the 
situation is similar. In this case the Fermi momenta are $\mu_{d_{L}}=\mu-\nu+\mu_5\approx 
\mu_5$ and $\mu_{d_{R}}=\mu-\nu-\mu_5\approx -\mu_5$, and in the region $\mu\approx\nu$ the 
condensation of
$\overline{d}_{R}d_{L}$ pairs is possible, and CSB phase is realized as well. The larger 
$\mu_5$ or $\nu_5$, the stronger the effect. And it can be realized in the broader band when 
the cancellation of $\mu$ and $\nu$ is less pronounced. The band of CSB phase started at 
around 0.15 GeV, let us
for simplicity round up to 0.2 GeV. As a result, the CSB phase is possible if 
$|\alpha+|\mu-\nu||$ is larger than 0.2 GeV (here $\alpha$ is $\mu_5$ or $\nu_5$). 
And one can see in Fig. 4 (a), Fig. 3 (c) that it is indeed the case.

One could also notice that the behaviour of, for example, charged PC phase at rather large 
values of $\nu$ in the case of $(\mu,\nu)$-phase diagram at nonzero chiral isospin $\nu_5$ and/or
nonzero chiral $\mu_5$ chemical potentials is rather different. For example, if $\nu_5\neq0$
one could behold the charged PC phase in the region of $\mu\approx\nu_5$ and $\mu_5=0$ (see in Figs. 3 (b) and (c)).
This difference could also be qualitatively
understood. The Fermi momenta of $u_{R}$ and $d_{L}$ quarks in this region are 
$\mu_{u_{R}}=\mu+\nu-\nu_5-\mu_5\approx\nu$ and $\mu_{d_{L}}=\mu-\nu-\nu_5+\mu_5\approx-\nu$ 
(see Eq. (\ref{1000})), so if the value
of isospin imbalance $\nu$ is rather large, then Cooper pairs condensation of the
type $\overline{d}_{L}u_{R}$ is possible and leads to charged PC. In the case of
$\mu_5\neq0$ the situation is drastically different, since in the regime $\nu_5=0$, 
$\mu\approx\mu_5$ and
with condition that $\mu,\mu_5$ are rather large together with $\nu$, the Fermi momenta of 
$u_{R}$ and
$d_{L}$ quarks are $\mu_{u_{R}}=\mu+\nu-\mu_5>0$ and $\mu_{d_{L}}=\mu-\nu+\mu_5>0$ and 
charged pion condensation is completely impossible.

Now let us contemplate the different setup, namely the $(\nu, \nu_5)$-phase diagram and the way
how it is affected by chiral imbalance $\mu_5$ at $\mu=0$.  It is fairly easy to obtain this phase
diagram, applying the ${\cal D}_2$-duality mapping to the corresponding $(\mu,\nu)$-phase diagram
at the same value of $\mu_5$ and at $\nu_5=0$. For example, the $(\nu, \nu_5)$-phase diagram at
$\mu_5=0.4$ GeV and $\mu=0$ is depicted in Fig. 4 (b), where
the band of BSF phase appears round the line $\nu=\nu_5$ when these chemical potentials are large
enough, $>$0.2 GeV. The diagram of Fig. 4 (b) is indeed a
${\cal D}_2$-duality mapping of Fig. 4 (a). Moreover, the BSF band {\it suddenly} appears on
these $(\nu, \nu_5)$-phase diagrams with $\mu=0$ even at smaller values of
$\mu_5>\mu_{5c}\in 0.15\div0.2$ GeV.
 So in this case the gears are switched and chiral
imbalance $\mu_5$ generates diquark condensation instead of chiral symmetry breaking.
Exactly the same band form of the BSF phase can be observed at the $(\nu, \nu_5)$-phase diagram with $\mu_5=0$ at 
different fixed
nonzero values of baryon (quark number) chemical potential $\mu$. One could easily envisage it by applying the
${\cal D}_2$-duality mapping to Fig. 3 (c)
and one can see that the behavior of these band-shaped regions, for $\mu\neq0$, $\mu_5=0$ and $\mu=0$, $\mu_5\neq0$, 
are exactly the same.  This is rather curious feature
that BSF phase is influenced exactly in the same fashion by baryon density and chiral density 
($\mu_5\ne 0$ or $\nu_5\ne 0$). As for the baryon density, $\mu\ne 0$, it is expected and natural to induce diquark
condensation, it is small wonder that at some regime the chemical potential $\mu$ leads to
diquark condensation, it does even in the simplest case of only $\mu\neq0$, but the fact that
chiral chemical potential $\mu_5$ is doing the same job is surprising enough. Even more surprising
that it does it in exactly the same way as baryon one, there is no difference whether there is baryon density in the system (with isospin and chiral isospin imbalance) or chiral imbalance.

Now let us reflect on the mechanism of emergence of the band-shaped region of BSF phase in 
$(\nu, \nu_5)$-diagrams of 
Fig. 4 (b) type that elongate along the region of $\nu_5\approx\nu$. In the most general case 
when all four chemical 
potentials are nonzero, the Fermi energies in this case are $\mu_{u_{R}}\approx\mu-\mu_5$, 
$\mu_{d_{R}}\approx\mu-\mu_5$ (it follows from Eqs. (\ref{1000}) at $\nu_5\approx\nu$). Hence, 
if $\mu_5=0$ and 
$\mu\neq0$ (as in the ${\cal D}_2$-duality mapping of the Fig. 3 (c) phase portrait) then the 
creation of Copper pairs of the 
form $u_{R}d_{R}$ is quite possible, and their condensation leads to the appearance of the 
band-shaped BSF phase along the line $\nu_5\approx\nu$ of the $(\nu, \nu_5)$-phase diagram at 
some fixed $\mu$. 
But in the case of the $(\nu, \nu_5)$-phase diagram of Fig. 4 (b) type, when $\mu=0$ and 
$\mu_5\neq0$ 
(for example $\mu_5=0.4$ GeV in Fig. 4 (b)), one can witness that the creation of Copper pairs 
of the form 
$\overline{u}_{R}\overline{d}_{R}$ is possible at $\nu_5\approx\nu$, also leading to the 
strip of the BSF phase in Fig. 4 (b).

Now consider the influence of chiral chemical potential $\mu_5$ on the $(\nu_5,\mu)$-phase diagram with some fixed 
$\nu$ and $\mu_5$, i.e. on the dense quark matter with chiral isospin imbalance.
The particular diagram of this kind is presented in Fig. 4 (c) at $\nu=0$ and $\mu_5=0.4$ GeV. And it is obtained 
(without any numerical calculations) simply by the action, e.g., of the ${\cal D}_3$-dual mapping on the diagram of 
Fig. 4 (a).
It is clearly seen that in Fig. 4 (c) the band-like region of charged PC phase is spawned at $\mu\approx\nu_5$, 
and this phase is ${\cal D}_3$-dually conjugated to the band-like CSB phase of Fig. 4 (a). Interestingly enough that 
this phase resembles the charged PC phase at the $(\mu, \nu_5)$-phase diagram at $\mu_5=0$ and $\nu=0.4$
GeV, which is a ${\cal D}_3$-dual mapping of the diagram in Fig. 3 (c). So in $\mu\approx\nu_5$ environment chiral 
imbalance $\mu_5$ fully takes a role of isospin $\nu$ one and causes the charged PC to crop up. In this region of
phase structure the effects of chiral imbalance $\mu_5$ and isospin imbalance $\nu$ is fully tantamount.

One can conclude that in the regime when values of two from basic triple of chemical potentials reach rather large 
values, the influence of chiral imbalance $\mu_5$ can be rather different, than just a universal catalyst for any 
pattern of symmetry breaking in the system. To support this statement, above we have considered in details the 
influence of $\mu_5$ on the phase structure of the model for three particular sets of basic chemical potentials, 
(i) $\nu_5=0$ but $\mu,\nu>0.2$ GeV, (ii) $\mu=0$ but $\nu,\nu_5>0.2$ GeV, (iii) $\nu=0$ but $\mu,\nu_5>0.2$ GeV. 
(In fact, the sets are dually conjugated to each other with respect to one of the dual transformations (\ref{55}) or 
(\ref{60}).) It turns out that at $\mu_5=0$ as well as at a rather small vicinity of the zero point,  the system is 
in the symmetrical phase for each set (i)-(iii) of basic chemical potentials (this conclusion is well illustrated by 
Fig. 3 (a) and its dual conjugations). However, when $\mu_5>\mu_{5c}$ 
(it is presented in Figs. 4 and considered in the above text of the present subsection), the chiral imbalance $\mu_5$ {\it induces} some 
symmetry breaking. In this case, depending on the conditions, $\mu_5$ can assume either the role of baryon 
chemical potential and generate the diquark condensation (for the set (ii)), or isospin chemical potential and 
trigger charged PC (for the set (iii)), or chiral isospin chemical potential $\nu_5$ and be a reason of CSB
(for the set (i)). And here it does not just catalyze the symmetry breaking and enhance the corresponding condensate 
as in chiral symmetry breaking catalysis observed when only $\mu_5\neq0$ \cite{braguta}, but {\it causes} the symmetry 
breaking to happen in the first place.

This leads to several curious consequences. For example, in the dense quark matter with rather considerable isospin 
imbalance when $\mu\approx\nu$, if one has nonzero chiral imbalance $\mu_5$ (exactly as if there was chiral 
imbalance $\nu_5$), then increasing the baryon density and keeping the ratio of baryon and isospin chemical potentials
the same, one could not restore chiral symmetry even at very large baryon density. Or, there is another one, 
if there is large baryon density and chiral isospin imbalance and $\mu\approx\nu_5$, then charged PC can be evoked in 
the system equally easily by isospin imbalance, $\nu\ne 0$, or by chiral imbalance, $\mu_5\ne 0$, and this would 
happen at any baryon density. %interestingly enough that the PC phase would be identical either it is $\mu_5\neq0$ or $\nu\neq0$.
One can say in this sense that the statement that the chiral imbalance $\mu_5$ has a chameleon nature, i.e. 
it can pretend to be either chemical potential depending on the situation, i.e. on the "colors of the environment", 
is supported once again. And it takes place in the case of more nonzero basic triple of chemical potentials and also, 
though in different hypostasis, in both regimes as for small and moderate values of these chemical potentials, 
as well as for rather large ones.

Let us reflect on the behaviour of chiral imbalance $\mu_5$ and its ability to generate various phenomena. It has been
shown that dualities (\ref{55}) and (\ref{60}) hold in the case of nonzero $\mu_5$ and they have powerful effect on 
the phase structure. If one knows that at $\mu_5>\mu_{5c}$ chiral imbalance generates chiral symmetry breaking in 
dense quark matter with isospin imbalance, $\mu\approx\nu$, see Fig. 4 (a), then one could use the duality and find 
out that chiral imbalance can generate diquark condensation and charged pion condensation, see Fig. 4 (b) and 
Fig. 4 (c).
So the feature of chiral imbalance $\mu_5$ of having the chameleon property is a consequence of dual properties of 
the phase diagram. $(\mu, \nu, \nu_5)$-phase diagram is highly symmetric and the expansion of this phase diagram to 
nonzero chiral chemical potential $\mu_5$ is highly symmetric, as well. So the influence of $\mu_5$ 
on the ($\mu$,$\nu$,$\nu_5$)-phase structure is constrained by duality and the chameleon nature of chiral chemical 
potential is inevitable. Or in other words, the full $(\mu, \nu, \nu_5,\mu_5)$-phase diagram at nonzero $\mu_5$ is 
highly symmetric due to dualities, and that chiral chemical potential $\mu_5$ does not participate in the duality 
transformations only keep the dualities of ($\mu$,$\nu$,$\nu_5$)-phase structure intact, all these leads to the fact 
that one can shuffle the facets of the ($\mu$,$\nu$,$\nu_5$)-phase diagram and due to this the influence of $\mu_5$ 
on various phenomena would be the same, since these facets are phases connected by duality transformations.
\label{IVC2}

\subsubsection{Dense quark matter with isospin and both chiral imbalances: the case of nonzero $\mu$, $\nu$, $\nu_5$, and $\mu_5$}

Now let us discuss the most generic situation when besides nonzero chiral imbalance $\mu_5$ all three basic chemical 
potentials are nonzero. This kind of general situation has been already considered in the regime of small and moderate
values of basic chemical potentials in Sec. IV B.
In this section, let us turn our attention to the regime where basic chemical potentials are rather large.

To begin with, we are going to investigate what happens to the $(\mu, \nu)$-phase diagram at nonzero fixed values 
of $\nu_5$, if one starts to increase the value of chiral chemical potential $\mu_5$. In particular, the 
$(\mu, \nu)$-phase diagram in the cases with only one chiral imbalance, i.e. at different nonzero values of $\mu_5$ 
and zero $\nu_5=0$ and at $\nu_5\neq0$ and $\mu_5=0$,  has been already deliberated in detail in Sec. \ref{IVC2}.
So first of all let us recall how the variation of chiral isospin $\nu_5$ chemical potential value would be imprinted 
at the $(\mu, \nu)$-phase diagram in the case $\mu_5=0$. Several diagrams of this type are shown in Fig. 3. As it is 
clear from it (see also in Ref. \cite{2color}) and has been discussed above, the boot-shaped elongated regions of 
charged PC and BSF phases at $\mu\approx\nu_5$ and $\nu\approx\nu_5$ are shifted to the larger values of $\mu$ and 
$\nu$, respectively, if one increases $\nu_5$. This effect, for example, leads to the generation of charged PC phase 
at larger baryon density, the similar effect in three color case led to the generation of charged pion condensation 
in dense quark matter with chiral imbalance \cite{kkz}.

Now let us return to the discussion of the influence of chiral chemical potential $\mu_5$ and recall in the first 
place that in the case of all four nonzero chemical potentials there is no symmetry of the TDP (\ref{72}) of the form 
$\mu\to-\mu$, etc. and, for example, the cases $\mu_5>0$ and $\mu_5<0$ has to be considered separately. So, in the 
most general case, one can consider $\mu>0$, $\nu>0$ and $\nu_5>0$ but the value of chiral imbalance $\mu_5$ could be 
of any sign or one can choose to consider $\mu>0$, $\nu>0$ and $\mu_5>0$ and the sign of $\nu_5$ is not fixed or any 
other combination (see the remark in the second paragraph below Eq. (\ref{72})).

Let us choose the former option, the one where basic chemical potentials $\mu$, $\nu$ and $\nu_5$ are positive, and 
first take the case of negative values of $\mu_5$. We would like to explore what effect chiral imbalance $\mu_5$ 
leaves on the $(\mu, \nu)$ diagram  at fixed $\nu_5\neq0$.
For example, to understand what would happen to Fig. 3 (c) if one increases the value of chiral imbalance $|\mu_5|$ 
provided that $\mu_5<0$, take a look at Fig. 6 (b). There the influence of the $\mu_5=-0.3$ GeV (instead of $\mu_5=0$
as in Fig. 3 (c))
on the $(\mu, \nu)$-phase diagram at fixed $\nu_5=0.4$ GeV is shown. As a result, one can see that the boot-like 
charged PC and BSF phases are shifted to the smaller values of $\nu$ and $\mu$ respectively, i.e. along the direction 
that they are elongated along (note that in the particular case of Fig. 6 (b) the shift is equal just to $\mu_5=-0.3$ 
GeV). Namely, charged PC and BSF phase shift by the quantity equal to around $\mu_5$ in the direction parallel to 
$\nu$ and $\mu$ axes, respectively. Let us note that it is the perpendicular direction to the one they shift when one 
increases $\nu_5$.

And at some rather large $|\mu_5|$ the charged PC and BSF phases of $(\mu, \nu)$-phase diagram with some fixed $\nu_5$
transfix the axes $\mu$ and $\nu$, respectively. Hence, a rather interesting picture unfolds, i.e. one could see that 
at large enough values of chiral imbalances $\nu_5$ and negative $\mu_5$ the diquark condensation takes place in the 
large values of $\nu$ and rather small $\mu$ region of phase diagram, but charged pion condensation is favoured in the 
region of large baryon (quark number) density and small isospin chemical potential $\nu$. 
\footnote{All this is in drastic contrast to the regular behavior at zero chiral imbalances $\nu_5$ and $\mu_5$ 
\cite{kogut,son2,weise,ramos, andersen3,brauner1,andersen2,imai,adhikari,chao,Bornyakov:2020kyz,khunjua,Furusawa:2020qdz}, 
or if only one of them is present, where BSF phase (charged PC phase) is generated when there is nonzero $\mu$ 
(isospin imbalance $\nu$), or if baryon chemical potential $\mu$ (isospin chemical potential $\nu$) has a higher 
values than the other ones \cite{2color}.} Especially interesting is the feature of dense quark matter when at nonzero
chiral imbalances, $\nu_5> 0$ and $\mu_5< 0$, and zero isospin one, $\nu=0$, the baryon chemical potential $\mu$
begins to increase 
(see, e.g., the movement along the line $\nu=0$ in Fig. 6 (b)). % if one consider the development with respect to changing baryon density. Let us stress that isospin imbalance is zero.
First, if $\mu$ is small or zero then CSB phase is realized due to rather large chiral imbalance $\nu_5$, then at 
larger baryon density the charged PC is triggered instead of diquark condensation. Due to a ${\cal D}_1$-self-duality 
of the $(\mu,\nu)$-phase diagram of Fig. 6 (b), a similar, but opposite, phenomenon occurs when at zero baryon 
chemical potential $\mu$ one increases the isospin one. It is clear from Fig. 6 (b) that at moderate values of $\nu$ 
less than 0.2 GeV chiral symmetry breaking predominates, due to chiral isospin imbalance $\nu_5>0$, but at larger 
values of $\nu$, instead of pion condensation that is usually promoted by isospin chemical potential, the diquark 
condensation develops in the system.
The remarkable feature is that even at exactly zero baryon chemical potential the diquark condensation is prevailing 
in full swing. Let us note here that in the BSF phase, which is located near the $\mu$-axis of the diagram in 
Fig. 6 (b) at nonzero values of isospin imbalance $\nu$, if one increases baryon chemical potential $\mu$ 
then diquark condensate grows.

Now let us consider the case $\mu_5>0$. Then the boot-like charged PC and BSF phases of the
$(\mu,\nu)$-phase diagram (at some fixed $\nu_5$ and $\mu_5$) are shifted to opposite direction
(compared to the case $\mu_5<0$), to the larger values of $\nu$ and $\mu$, respectively, when
$\mu_5$ increases. For example, in Fig. 6 (a) the $(\mu,\nu)$-phase diagram in the case
$\nu_5=0.25$ GeV and $\mu_5=0.5$ GeV is depicted. The value of $\nu_5=0.25$ GeV is taken for a
change, and one can easily envision the $(\mu,\nu)$-phase structure at $\nu_5=0.4$ GeV and
$\mu_5=0.5$ GeV, where the charged PC and BSF phases would be shifted to the larger values of
$\nu$ and $\mu$, respectively, in comparison with the $(\mu,\nu)$-phase diagram of Fig. 3 (c)
at $\nu_5=0.4$ GeV and $\mu_5=0$.
We will discuss the additional reasoning apart from diversity and clarity for this choice below,
now let us turn back to phase structure of Fig. 6 (a). While these boot-like phases are moved to
the opposite direction, i.e. larger values of $\nu$ and $\mu$ chemical potentials, there a new
boot-like region of BSF (charged PC) phase appears from the small values of $\nu$ ($\mu$).
To demonstrate in more details this effect, in Fig. 2 (b) the behaviour with respect to $\nu$ at
$\mu=0$ and orders of magnitudes of the condensates are depicted. One could note from this figure
that a rather large diquark condensate, larger than around 0.15 GeV, could be generated in quark
matter even at zero baryon density.

Thus, the analysis of the phase diagrams in Fig. 6 allows us to draw a very interesting conclusion
that even at zero baryon density, $\mu=0$, in quark matter with an increase of $\nu$, a BSF phase
can arise, which is {\it induced} in the system only in the presence of a rather large chiral
asymmetry ( large value of $|\mu_5|$). From the diagrams in Fig. 6, a dually-${\cal D}_1$
conjugate conclusion also follows: if the isospin density in quark matter is zero, $\nu=0$, then
with increasing of $\mu$ in it, in the presence of a chiral asymmetry with a large value of
$|\mu_5|$, a charged PC phase is {\it generated}. These phenomena are quite curious and they could not be
observed in the system without chiral imbalance. We especially emphasize that these phenomena are
precisely {\it generated} by the chemical potential $\mu_5$, since at $\mu_5=0$ and at the same
values of the basic chemical potentials, neither the BSF nor the charged PC phase
is present in the system (see Fig. 3 from Ref. \cite{2color}).

In part, the phase structure of the diagrams in Fig. 6 can also be explained qualitatively
at the level of Fermi energies (chemical potentials) for left- and right-handed quarks, 
by employing the pairing arguments. In the region of, for example, charged PC phases of Fig. 6,
where $\mu\approx\nu_5$, we have from Eq. (\ref{1000}) that $\mu_{u_{R}}\approx\nu-\mu_5$ and 
$\mu_{d_{L}}\approx-\nu+\mu_5$.
So one can see that if $\nu=0$ (or rather small value), then $\mu_5\neq0$, both negative and
positive, could generate charged PC phase, as it happens in all diagrams of Fig. 6. Now 
suppose that
$\nu$ is quite large, $\nu>0.5$ GeV, and still $\mu\approx\nu_5$. Then in Fig. 6 (a), at 
$\mu_5>0$,
you can see the charged PC phase (since in this case there is a condensation of the 
$\overline{d}_L u_R$ Cooper pairs)
and in Fig. 6 (b), at $\mu_5<0$, there is no. We explain the
difference as follows. In this case, at $\mu_5<0$ the quasiparticles $\overline{d}_{L}$ and 
$u_{R}$
will have a very large relative momentum, $\sim (\nu+|\mu_5|)$, and they will not have enough time
to create a Cooper pair. Therefore, there is no charged pion condensation, and in the
corresponding region of Fig. 6 (b) the charged PC phase is absent. But in the case when
$\mu_5>0$, their relative momentum is rather small, $\sim (\nu-\mu_5)$, and the quasiparticles
have enough time to form a Cooper pair, the condensation of which leads to the charged PC phase
as in Fig. 6 (a). \footnote{In a similar way, it is possible to explain qualitatively the presence of the BSF 
phase in Fig. 6, etc.}
\begin{figure}
%----figure 6
\hspace{-1cm}\includegraphics[width=1.0\textwidth]{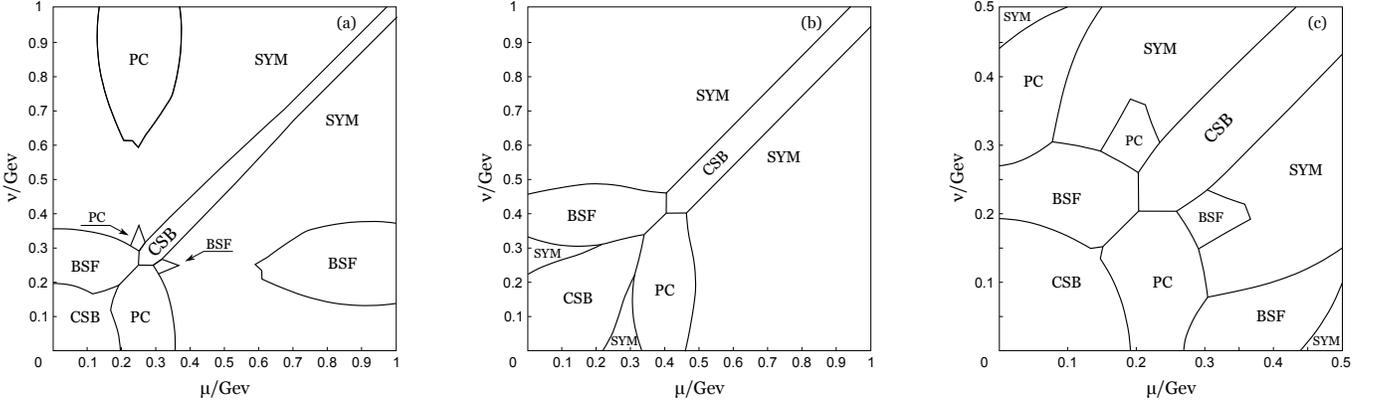}
% \hfill
%\includegraphics[width=0.33\textwidth]{munu_nu5_04_mu5_m03.eps}
%\hfill
%\includegraphics[width=0.33\textwidth]%%%{munu_nu5_02_mu5_05_l.eps}\\
%\parbox[t]{0.45\textwidth}{
 \caption{ $(\mu,\nu)$-phase diagrams: (a) at $\nu_{5}=0.25$ GeV, $\mu_5=0.5$ GeV,
 (b) at $\nu_{5}=0.4$ GeV, $\mu_5=-0.3$ GeV, (c) at $\nu_{5}=0.2$ GeV, $\mu_5=0.5$ GeV.
 The notations are from Fig. 1.}
\end{figure}

Let us now elaborate on the choice of $\nu_5=0.25$ GeV in Fig. 6 (a). The pattern of the phase
structure that we have talked about above was inherent only in the regime of rather large values
of chiral isospin imbalance $\nu_5$, one can say that larger than $0.2$ GeV, otherwise the
boot-like regions would not emerge and separate from the $\nu$ and $\mu$ axis. (The regime of
values smaller than $0.2$ GeV has been discussed in Sec. IV B.) One can see in Fig. 3 (b) that these
phases just start to appear at $\nu_5=0.2$ GeV. So for $\nu_5=0.25$ GeV the discussed regime works
in full fledged mode. Let us consider the transitionary regime between moderate values of $\nu_5<0.2$ GeV
and rather large ones $\nu_5>0.2$ GeV and see how the ($\mu, \nu$)-phase diagram depicted in Fig. 3 (b) alters,
if the chiral imbalance $\mu_5$ is increased. Although at $\nu_5=0.2$ GeV and $\mu_5=0$ the
regime of moderate values still worked, see Fig. 3 (b), when one increases $\mu_5$ to rather large
values, the rather simple picture breaks down and one can see in Fig. 6 (c), where the
($\mu, \nu$)-phase diagram at $\mu_5=0.5$ GeV and $\nu_5=0.2$ GeV is depicted, that the phase
structure is quite rich and complicated. Indeed, there the new regions of charged PC and BSF phases at $\mu=\nu_5$ 
and $\nu=\nu_5$ appear, as in the case of regime of large $\nu_5$, but at larger
values of $\mu$ and $\nu$, respectively. The more expected BSF and charged PC phases emerge,
as in the case of moderate values. The phase structure is rather complicated, for example, for 
small or zero values of isospin imbalance $\nu$. If in this case the baryon density, or $\mu$, 
is increased from small values, first CSB phase takes place and it is not a surprise, since chiral 
imbalance is rather large in the system. Then, if baryon density is increased, charged pion 
condensation is {\it triggered} in the system. Comparing Fig. 6 (c) with a similar phase diagram 
of Fig. 3 (b) with the same fixed $\nu_5=0.2$ GeV but $\mu_5=0$, we see that this effect is just 
due to a nonzero and large value of $\mu_5$. Moreover, one could note that in this case the 
generated value of the charged pion condensate $\pi_1$ is rather large, see in Fig. 2 (c). And then, 
at still larger values of baryon density, diquark condensation, or BSF phase, is generated. 
In this case the phase diagram is quite involved and rich, and a lot of first order phase 
transitions can happen in the range of baryon chemical potential $\mu$ from zero, or even around 
0.2 GeV, to $0.3$ GeV. By a ${\cal D}_1$-self-duality of the ($\mu, \nu$)-phase diagram of Fig. 6 (c), 
it is clear that at zero or small values of $\mu$ the isospin chemical potential $\nu$ is able to generate the 
BSF phase at nonzero values of $\mu_5$.

One can also note that changing the values of chiral imbalances $\mu_5$ and $\nu_5$ one can get
the boot-like BSF and charged PC phases in any part of the ($\mu$, $\nu$)-phase diagram
(compare the diagrams in Fig. 6 and Fig. 3). Moreover, applying to the diagrams of Fig. 6 the ${\cal D}_2$- and
${\cal D}_3$-dual transformations, it is possible to obtain the $(\nu, \nu_5)$- and
$(\mu, \nu_5)$-phase diagrams of the model at the same fixed values of $\mu_5$. \vspace{0.15cm}

Let us summarize in a way and ingeminate the following points. In the regime when basic chemical 
potentials have small or moderate values, and at least two of them are less than $0.2$ GeV, chiral 
chemical potential $\mu_5$ could only catalyze the phases triggered by basic triple of chemical 
potentials. In this case it is a universal catalyst, since it catalyzes all the symmetry breaking 
patterns in the system. But in the regime when basic chemical potentials are rather large, and at least 
two of them are larger than or around $0.2$ GeV, chiral chemical potential $\mu_5$ can entail rather peculiar
symmetry breaking patterns. For example, one can note that charged PC can emerge in the system with
zero isospin chemical potential. Or at $\mu_5\ne 0$ the diquark condensation can be realized 
in quark matter with isospin asymmetry, $\nu\ne 0$, but zero baryon density, $\mu=0$ (see in Fig. 6).

The feature of chiral imbalance $\mu_5$ of having the chameleon property is a consequence of dual
symmetries of the full phase diagram of the model. 
%($\mu$, $\nu$, $\nu_5$)-phase diagram is highly symmetric due to the dualities and the expansion 
%of this phase diagram to non-zero chiral chemical potential is also highly symmetric. 
%For example, if one imagine ($\mu$, $\nu$, $\nu_5$)-phase diagram as a cube, then one can pick 
%three facets of this phase diagram, namely ($\nu_5=0$ and $\nu=0$), ($\nu=0$ and $\mu=0$) and 
%($\mu=0$ and $\nu_5=0$) then these facets are ($\mu$, $\mu_5$), ($\nu_5$, $\mu_5$) and ($\nu$, 
%$\mu_5$). So due to reservation retention of dualities at $\mu_5\neq0$ these facets are dual to 
%each other. 
Hence, if, e.g., the CSB phase is catalysed, as in Fig. 1 (a), by chiral imbalance $\mu_5$ at 
$\nu_5\neq0$, then the ${\cal D}_3$-dual symmetry of the phase diagram  
dictates that chiral imbalance $\mu_5$ catalyses the charged pion condensation at 
$\nu\neq0$, see Fig. 1 (b), and that it catalyses the diquark condensation at $\mu\neq0$, 
see Fig. 1 (c), due to the ${\cal D}_3$-dual symmetry. More generally, if chiral imbalance $\mu_5$ 
provokes some phenomenon in some 
cross-section of the full phase diagram, then the dually conjugated phenomenon is triggered by $\mu_5$ in 
the dually conjugated cross-section of the full phase diagram. So the chameleon nature of chiral chemical
potential $\mu_5$ is inevitable consequence of dualities.

\section{Summary and Conclusions}
In this paper the influence of chiral imbalance $\mu_5$ on such phenomena of two-color quark matter as chiral 
symmetry breaking, diquark and charged
pion condensations has been scrutinized in the framework of two-color effective
NJL model in the mean-field approximation. The influence of other, quark number (or baryon) $\mu$, isospin $\nu$ and 
chiral isospin $\nu_5$, chemical potentials has been investigated in our previous article \cite{2color}. The main 
results of the paper are the following:
%\begin{itemize}
 
$\bullet$ It was shown in the mean-field approximation that at $\mu_5\ne 0$
the thermodynamic potential  (\ref{72}) of the model has the same three
dual symmetries ${\cal D}_1$ (\ref{55}), ${\cal D}_2$ and ${\cal D}_3$ (\ref{60}),
found in the case of $\mu_5=0$. These dualities lead to some fundamental discrete symmetries of the full  
$(\mu,\nu, \nu_5,\mu_5)$-phase diagram of the model, which we also call dual symmetries.
 
$\bullet$ Since the duality transformations
${\cal D}_1$, ${\cal D}_2$ and ${\cal D}_3$ do not involve $\mu_5$, chiral $\mu_5$ stands alone from other chemical 
potentials,
baryon, isospin and  chiral isospin chemical potentials,  as we call them basic ones,
in the sense that at fixed $\mu_5$ all other  chemical potentials are intermingled together by
dualities and
amalgamated in some  $(\mu, \nu, \nu_5)$ cross-section of the full phase diagram of the model. Due to this fact
 %the fact 
 %that dualities act only in the 
 %sections at fixed $\mu_5$  
the full $(\mu,\nu, \nu_5,\mu_5)$-phase diagram could be envisaged as a foliation of the 
$(\mu,\nu, \nu_5)$-phase portraits along the axis of chiral imbalance $\mu_5$. And each $\mu_5\ne 0$-fixed 
cross-section of the full phase diagram, or its  $(\mu,\nu, \nu_5)$-phase portrait, is highly symmetric to the same
extent as at $\mu_5=0$.
 
$\bullet$ This fact %that at fixed $\mu_5\ne 0$ the phase diagram $(\mu, \nu, \nu_5)$ is highly symmetric 
entails the number of very interesting implications for the phase structure. It will be discussed later but let us 
recall here that due to it various cross-sections of $(\mu, \nu, \nu_5)$-phase diagram are self-dual 
(for example, each $(\mu, \nu)$-phase diagram of Figs. 3 and 6 is a ${\cal D}_1$-self-dual). So it is very easy 
 to analyze it by performing numerical calculations of one of its phase portraits, and then applying to it different 
 duality mappings (e.g., in Fig. 4 the middle and right panels are, respectively, ${\cal D}_2$ and ${\cal D}_3$ 
 conjugated to Fig. 4 (a)). If one varies the chiral imbalance $\mu_5$, then the $(\mu, \nu, \nu_5)$-phase diagram 
 as a whole is distorted and transformed, but its property of high symmetry remains intact. So if one fixes the value 
 of $\mu_5\neq0$, then this phase diagram could be explored with ease, since one need only limited number of its 
 cross-sections to reproduce the whole $(\mu, \nu, \nu_5)$-phase diagram due to these high symmetry feature.

 $\bullet$ Considering the particular scenario when the system possesses chiral imbalance $\mu_5$ and only one 
 of basic chemical potentials is nonzero (see in Sec. IV A), one can already descry interesting features of 
 chiral chemical potential $\mu_5$.

 First, let us note that nonzero chiral imbalance $\mu_5$ does not handicap the property of the phase structure that 
 each basic chemical potential is connected by one-to-one correspondence to some phenomenon that it generates, i.e.
 chiral isospin chemical potential $\nu_5$ generates CSB, isospin chemical potential $\nu$ entails charged PC 
 phenomenon and $\mu$ engenders BSF -- diquark condensation phenomenon.

Second, in this scenario chiral chemical potential $\mu_5$ takes a role of the nonzero basic chemical potential
and catalyzes the corresponding phenomenon. (In this case the corresponding condensate increases
vs $\mu_5$, see, e.g., the behavior of $M$ vs $\mu_5$ in Fig. 2 (a)). In other words, it acts as
a phantom of this basic chemical potential.
More specifically, $\mu_5$ can play a role of chiral isospin chemical
potential $\nu_5$, if $\nu_5\neq0$, and catalyze CSB phenomenon. It can also take a role of isospin chemical 
potential $\nu$, if $\nu\neq0$, and catalyze charged PC, or equally easy it can take the role 
%assume the role be a double be a counterpart
of baryon chemical potential $\mu$, if $\mu\neq0$, and enhance diquark condensation. Taking on the role of any basic 
chemical potential, if it is nonzero, can be compared to a chameleon property to assume the colour of the environment
in which it is placed. If there is only chiral isospin chemical potential $\nu_5$ in the system, then $\mu_5$ pretends
to be $\nu_5$ and catalyzes CSB phenomenon. If there is baryon density $\mu$, then chiral chemical potential $\mu_5$ 
mimics and enhances its ability to promote the BSF phenomenon, etc.

The catalysis of chiral symmetry breaking by $\mu_5$ found earlier in \cite{braguta} is only one facet 
(particular case) of the general picture of capacity of chiral imbalance $\mu_5$ to catalyze various phenomena. 
This particular case is taking place if there is no other basic chemical potentials to trigger other phases in the 
system, rather than the phase with chiral symmetry breaking. The $\mu_5$ catalysis effect of any phenomena is a 
rather universal feature of chiral imbalance, $\mu_5$ catalyzes any phenomena triggered in the system by a 
corresponding single nonzero basic chemical potential and does not prefer any of them.

Rather curious feature is that chiral imbalance $\mu_5$ needs only a seed (in the form of only one
nonzero basic chemical potential) to trigger the corresponding condensation phenomenon. Then it
could fully replace the corresponding basic chemical potential and enhance this phenomenon on its
own. For example, if $\mu\neq0$ and is rather small, $<\mu_5$, then $\mu_5$ can enhance diquark
condensation to the levels, observed at similar values of $\mu\approx\mu_5$. Diquark condensate
would have the same value in the following cases, (i) if $\mu$ is rather large and $\mu_5$ is
small and, a reversed one, (ii) if $\mu\ll\mu_5$. In both cases, (i) and (ii), the diquark
condensate $|\Delta|$ takes on the value it would have if there were only one baryon chemical
potential in the system equal to ${\rm max}\{\mu_5,\mu\}$.

 %\item
$\bullet$ The full $(\mu,\nu, \nu_5,\mu_5)$-phase structure of the model could be divided into two regimes, 
in the first one, when $\mu_5\ne 0$ and basic chemical potentials have small or moderate
values $<0.2$ GeV, the picture is quite frugal in details, but concise and elegant (see in Sec. IV B). Indeed, to 
find the phase structure in this
case, even at $\mu_5=0$, one should perform several simple steps. (i) First, it is necessary to
connect  each  possible phenomenon of two-color quark matter with one of the basic chemical
potentials according to a rule: CSB$\leftrightarrow\nu_5$, charged PC$\leftrightarrow\nu$ and
BSF$\leftrightarrow\mu$.
%It is necessary to put each possible phenomenon of two-color quark matter in a one-to-one correspondence with one of the basic chemical potentials according to a rule: CSB$\leftrightarrow\nu_5$, charged PC$\leftrightarrow\nu$ and  BSF$\leftrightarrow\mu$.
(ii) Then it is necessary to find the biggest one among all basic chemical potentials.
And (iii) this (biggest) chemical potential corresponds according to a rule (i) to a phase,
which is realized in the system in this case.
(A special simple case of this scheme has been discussed in Sec. IV A,
where the case with only one nonzero basic chemical potential has been considered.)
Hence, in this case the whole phase diagram is quite easy to comprehend, it can be sketched in the
few lines. The relation between three basic chemical potentials fully defines the phase structure
in the rather concise fashion, the largest chemical potential triggers the corresponding symmetry
breaking pattern and hence the corresponding phenomenon.
A quite eye-catching effect is that while increasing the other two basic chemical potentials up
to the point, where they are less than the largest one, they do not have almost any impact on
the phase structure and can be disregarded in a way.

 %  \item
%$\bullet$
The chiral chemical potential $\mu_5$ stands alone from the basic chemical potentials. Its
influence on the phase diagram in this regime is drastically different to other (basic)
chemical potentials. While the relation between values of basic chemical potentials determine the phase settled in
the system, the increase of chiral chemical potential $\mu_5$ does not meddle in this phase
rivalry. It catalyzes any of these phases picked by basic chemical potentials and drastically enhances the value of 
the corresponding condensate. The chameleon nature of $\mu_5$ is demonstrated in a full swing here, chiral chemical
potential assumes the property of the largest basic chemical potential and catalyzes the
corresponding phenomenon.

%  \item
$\bullet$ In the second regime of rather large values of basic chemical potentials, at least when two
of them are larger than 0.2 GeV, the phase diagram is much more involved and rich and not that
concise in the form, especially if there are all three nonzero basic chemical potentials, see, e.g.,
in Fig. 6.

%  \item
%$\bullet$
In this regime, already in the case of two nonzero basic chemical potentials,
the chiral imbalance $\mu_5$ is not just a universal catalyst for any pattern of symmetry breaking
in the system. It can also induce some breaking on its own, i.e. at zero values of $\mu_5$, the
system would be in symmetric phase but if $\mu_5\neq0$ some symmetry ends up being broken. But depending on the
conditions, $\mu_5$ can assume either the
role of baryon chemical potential $\mu$ and generate the diquark condensation, or isospin chemical
potential $\nu$ and trigger charged PC, or chiral isospin chemical potential $\nu_5$ and be a
reason of chiral symmetry breaking.
In this case as well chiral chemical potential $\mu_5$ can pretend to be any basic chemical
potential depending on the situation, and in this sense the statement that the chiral imbalance
has a chameleon nature is supported once again.

% \item
$\bullet$ Chameleon nature of chiral imbalance is rather universal phenomenon and it can be
observed in both regimes, as for small and moderate values of basic chemical potentials,
as well as for rather large ones, though in different hypostasis. In the first case $\mu_5$
{\it catalyses/enhances} some phenomenon, in the other it could {\it induce} it on its own. But in both cases
it mimics the corresponding basic chemical potential. Chiral imbalance $\mu_5$ can promote all
phenomena happening in two-color quark matter.  This property of chiral imbalance is based
on the fact that at fixed $\mu_5$ the $(\mu, \nu, \nu_5)$-phase diagram of the two-color NJL model
is highly symmetric due to dualities (see in Sec. III).
All these leads to the fact that, using duality mappings, one can shuffle the facets of the
$(\mu, \nu, \nu_5)$-phase diagram. So if $\mu_5$ promotes some phenomenon, it could promote
two others in the dually conjugated sectors of the $(\mu, \nu, \nu_5)$-phase diagram.
So the influence of chiral chemical potential $\mu_5$ on the phase structure  is highly
constrained by dualities, its chameleon nature is inevitable and it is a consequence of duality
properties.\vspace{0.15cm}

In conclusion, we would like to note one more interesting property of chiral imbalances in two-color quark matter. 
Suppose that it is somehow possible to obtain in the system nonzero values of the chiral density of both types and, 
moreover, to change the values of the corresponding chemical potentials, $\nu_5$ an $\mu_5$. Then practically for 
arbitrary 
fixed values of $\mu$ and $\nu$ it is possible to get both BSF and charged PC phases, selecting for each of these 
phenomena the corresponding values of $\nu_5$ and $\mu_5$. For example, it is clear from Fig. 6 that 
at large enough values of chiral imbalances $\nu_5$ and $\mu_5$ the diquark condensation takes place in the region of 
rather large values of isospin imbalance $\nu$ and small values of $\mu$. And charged PC is favoured in the region of 
large baryon $\mu$ and small isospin $\nu$ chemical potentials. This is in drastic contrast to the familiar and 
natural behavior observed, e.g., in Ref. \cite{andersen2} at zero chiral imbalance, where BSF phase (charged PC phase)
is generated when there is rather large baryon density (isospin imbalance).

The contemplated in the paper phase diagram shows the inherent elegance of the phase 
structure of two color QCD and elucidates properties of the chiral imbalance and its effect 
on the phase structure that could appear to be universal and inherent also to the three color 
QCD.

\section{ACKNOWLEDGMENTS}

R.N.Z. is grateful for support of Russian Science Foundation under the grant  19-72-00077. The work was also supported by the Foundation for the Advancement of Theoretical Physics and Mathematics BASIS.

\end{document}